\documentclass[aps,12pt,a4paper]{revtex4}
\usepackage{epsfig}
\usepackage{graphicx}
\usepackage{amsmath,amssymb,color}
\usepackage[english]{babel}

\parskip=\medskipamount

\newcommand{\eq}[1]{(\ref{#1})}
\newcommand{\fig}[1]{Fig.~\ref{#1}}

\newcommand{\be}{\begin{equation}}
\newcommand{\ee}{\end{equation}}
\newcommand{\beq}{\begin{equation}}
\newcommand{\eeq}{\end{equation}}
\newcommand\disp{\displaystyle}

\newcommand{\la}{\left<}
\newcommand{\ra}{\right>}
\newcommand{\eps}{\varepsilon}
\newcommand{\Tr}{\mathrm{Tr}\,}

\newcommand{\im}{\textrm{Im}\,}

\begin{document}

\title{Lifshitz tails at spectral edge and holography with a finite cutoff}

\author{Alexander Gorsky$^{1,2}$, Sergei Nechaev$^{3,4}$, and Alexander Valov$^{5}$}

\affiliation{$^1$ Institute for Information Transmission Problems RAS, 127051 Moscow, Russia \\ $^2$ Moscow Institute of Physics and Technology, Dolgoprudny 141700, Russia \\ $^3$ Interdisciplinary Scientific Center Poncelet (CNRS UMI 2615), 119002 Moscow, Russia \\ $^4$ P.N. Lebedev Physical Institute RAS, 119991 Moscow, Russia \\ $^5$ Physics Department, Lomonosov Moscow State University, 119992 Moscow, Russia}

\begin{abstract}

We propose the holographic description of the Lifshitz tail typical for one-particle spectral density of bounded disordered system in $D=1$ space. To this aim the "polymer representation" of the Jackiw-Teitelboim (JT) 2D dilaton gravity at a finite cutoff is used and the corresponding partition function is considered as the weighted sum over paths of fixed length in an external magnetic field. We identify the regime of small loops, responsible for emergence of a Lifshitz tail in the Gaussian disorder, and relate the strength of disorder to the boundary value of the dilaton. The geometry corresponding to the Poisson disorder in the boundary theory involves random paths fluctuating in the vicinity of the hard impenetrable cut-off disc in a 2D plane. It is shown that the ensemble of "stretched" paths evading the disc possesses the Kardar-Parisi-Zhang (KPZ) scaling for fluctuations, which is the key property that ensures the dual description of the Lifshitz tail in the spectral density for the Poisson disorder.

\end{abstract}

\maketitle

\tableofcontents

\section{Introduction}
\label{s:1}

Emergence of so-called "Lifshitz tail" (LT) in the eigenvalue density, $\rho(E)$, at the spectral edge of disordered systems \cite{lifshitz} is a well-known phenomenon in condensed matter physics, originating from the Griffiths-type rare events. Specifically, the Lifshitz tail manifests itself in the singular distribution of eigenvalues in the vicinity of the spectral edge of the Schr\"{o}dinger operator with the diagonal disorder. The Lifshitz singularity is ubiquitous for spectral densities of various one-dimensional disordered systems. To name but a few, we can mention random sequences of light and heavy masses connected by harmonic springs \cite{rodgers,neuwen1}, sparse random matrices \cite{khorun}, 1D systems with random hopping \cite{polov}, trapping problems in 1D disorder environment \cite{neuwen2}, some number-theoretic aspects of continued fractions \cite{borwein}. In Appendix \ref{app1} we discuss in more details some of these systems.

The eigenvalue density, $\rho(E)$, near the spectral edge depends on the type of the disorder. For the Gaussian disorder in the $D$-dimensional space, $\rho(E)$ behaves at $E\to -\infty$ as follows:
\be
\rho(E) \propto |E|^{\frac{D(5-D)}{4}} \exp\left(-\alpha |E|^{\frac{4-D}{2}}\right)
\label{e:001}
\ee
where $\alpha$ is some model-dependent constant. Thus, for $D=1$ one gets
\be
\rho(E) \propto |E| \exp\left(-\alpha |E|^{\frac{3}{2}}\right)
\label{e:001a}
\ee
On the other hand, for the Poissonian disorder in the $D$-dimensional space, the behavior of $\rho(E)$ at $E\rightarrow 0$ is different:
\be
\rho(E)\sim \exp\left(-c_D E^{-\frac{D}{2}}\right)
\label{e:002}
\ee
where $c_D$ is some constant which depends on the disorder strength. If the spectrum of a $D$-dimensional disordered system has a finite support $[-E_c,E_c]$, then the spectral density, $\rho(E)$, near the spectral edge $E_c$ has the following asymptotic form
\be
\rho(E) \propto \exp\left(-c_D |E-E_c|^{-\frac{D}{2}}\right)
\label{eq:n(e)}
\ee
providing the  LT singularity for 1D systems with the Poissonian disorder. Performing the Laplace transform of \eq{eq:n(e)}, we get
\begin{multline}
Z(\beta) \propto \int_0^{\infty} \exp\left(-c_D |E-E_c|^{-\frac{D}{2}}\right) e^{-\beta E} dE\,\bigg|_{\beta\gg 1} \propto \\ \beta^{-\frac{D+4}{2(D+2)}} \exp\left(-\beta E_c - b_D c_D^{\frac{2}{D+2}} \beta^{\frac{D}{D+2}}\right)
\label{eq:n(L)}
\end{multline}
where $b_D=\left(\frac{d}{2}\right)^{\frac{2}{D+2}}\left(1+\frac{2}{d}\right)$ is the $D$-dependent constant (for example, $b_1=3\times 2^{-2/3}$). The Griffiths-type sublinear ("stretched exponential") behavior $\sim \exp(-\mathrm{const}\;\beta^{D/(D+2)})$ in \eq{eq:n(L)} is well known in physics of disordered systems since it describes the survival probability of a diffusive particle trapping in the $D$-dimensional space with the Poissonian distribution of traps. Later on we relate the 1D behavior $\sim \exp(-\mathrm{const}\; \beta^{1/3})$ to the (1+1)D Kardar-Parisi-Zhang (KPZ) scaling exponent. In what follows it is useful to discretize the Euclidean time interval, representing $\beta$ as $\beta = aN$, where $a$ is the elementary time step.

In our study we are attempting to develop the holographic counterpart of the Lifshitz tail phenomenon. The standard viewpoint is such that the spectral edges are most suitable objects for the dual holographic description and the very question seems reasonable, though nontrivial. The holographic approach with the gravity in the bulk usually implies that one deals with large-$N$ and strong coupling  in the boundary theory. Below we are discussing the (0+1) boundary quantum mechanical system on an Euclidean time circle. There are illuminating examples of holography for (0+1) large-$N$ systems. First of all, the BFSS model \cite{bfss} should be mentioned in that context, which describes the system of large-$N$ number of D0-branes and has a classical gravity dual counterpart in some interval of parameters. Among more recent examples, the (0+1) SYK model with the Gaussian disorder \cite{sy,k} takes a special place, providing an at low energy important pattern for the holographic description with the $AdS_2$ background gravity and the dilaton field. The $AdS_2$ geometry plays a distinguished role in the holographic context since it describes the near-horizon region in higher dimensional RN-black holes (BH). The low-energy sector of SYK model \cite{sy,k} governed by the Schwartzian action, gets mapped onto the JT gravity \cite{j,t}.

In both models, degrees of freedom providing large-$N$ behavior interact with each other and the strong coupling between degrees of freedom is an important ingredient of the holographic picture. As it is shown below, in our problem there is a large number of defects with Gaussian or Poissonian distributions, wrapping around Euclidean time circle. Their presence indeed ensures the large-$N$ behavior in the boundary theory. We have no any gauged $SU(N)$ symmetry at the boundary, but it was argued in \cite{mm} that in (0+1)holography such a symmetry is not crucial. We perform ensemble averaging over defects distributed along $D$ space directions which holographically are target space coordinates, while in the SYK case there is no such distribution since all fermions are
sitting at one point. In BSSF model in principle such an averaging over the "Coulomb branch" distribution of positions of D0 branes can be considered.

Few years ago the JT gravity with the cut-off has been reformulated as a "polymer"  path integral  in the 2D $AdS_2$ hyperbolic geometry \cite{kitaev,yang}. More refined analysis \cite{stanford} has clarified the important role of the self-avoidness of particle's trajectories. The asymptotic value of the dilaton $p$ plays the role of a pressure, or of an external magnetic field, which affects the typical trajectory. It has been demonstrated that the partition function of the JT gravity with the cut-off, can be written as the sum over the ensemble of self-avoiding trajectories of fixed length in the hyperbolic plane in a constant transverse magnetic field. The length of the trajectory, $\beta=T^{-1}$, has a meaning of an inverse temperature, $T$, in the boundary theory. Different regions at $(\beta,p)$ plane correspond to different sub-families of trajectories and yield several overlapping regimes with specific tails of a spectral density. The JT gravity with the cutoff \cite{stanford,verlinde,hartmann} can be also interpreted in terms of the $T\bar{T}$ deformed boundary quantum mechanics and the deformation parameter is related to the cut-off scale \cite{gross}. The partition sum in the theory with cutoff has been identified with the solution to the WdW equation \cite{verlinde}.

Implying the holographic description of (0+1) system, describing the probe particle in the random environment, and keeping in mind all above mentioned reservations, we address the following questions:
\begin{itemize}
\item Which ensembles of trajectories in the "polymer" interpretation of JT gravity are relevant for emergence of the Lifshitz tail (if any)?
\item What is the holographic counterpart of a disorder strength?
\item Does the behavior of a holographic system in the bulk depend on the disorder type, namely, on Gaussian/Poisson distributions?
\end{itemize}

In the holographic context the relevant dual bulk coordinates of the boundary quantum mechanics are $(t_E,r,x_1\dots x_D)$ where we do not touch the rest coordinates in 10D geometry. In what follows we try to derive the LT from the holography in a 2D space involving the radial coordinate, $r$ and Euclidean time, $t_E$. There are two immediate obstacles for such a viewpoint. First, we know that the LT follows from the rare event scattering  of the probe particle at the defects in the $D$-dimensional space. How such physical process can be captured in the $(t_E,r)$ plane only? Secondly, the spectral density depends on the space dimensionality, $D$, hence naively we have to consider the whole $(t_E,r,x_1\dots x_D)$ bulk.

Hopefully the $D=1$ case appears to be special and both problems can be avoided. First, we use the instanton derivation of the Lifshitz tail which can be expected from its exponential behavior. It turns out that upon the disorder averaging, the theory admits description in terms of an effective action, $S$, of a specific sigma model \cite{cardy, halperin,langer}. The correct account of instanton-like solution to $S$ yields the requested exponential behavior in the spectral density both for the Gaussian and Poissonian disorders \cite{langer,luttinger,Yaida}. In \cite{lubensky, renn} the general effective action providing the "master" instanton solution for the LT has been proposed. The most important point of the instanton interpretation is that the LT is derived from the instanton solution of an effective sigma model strongly localized at some radial coordinate $r_0$ in $(x_1\dots x_D)$ space. Hence in $D=1$ the solution implies the localization at the point while for $D>1$ the localization takes place at the manifold $S^{D-1}$. Thus at $D=1$ the space dimension is frozen and we could expect the holographic description with $(t_E,r)$ bulk. That is  why we are mainly focused at the Lifshitz tail in the one-particle spectral density in  the $D=1$ disordered boundary theory. Like in \cite{stanford}, our main tool is the representation of the gravity path integral in terms of the quantum mechanics of a particle in an effective magnetic field. Using the large-deviation description, we identify the sub-ensemble of trajectories in JT gravity which is relevant for emergence of LT in the spectral density of the boundary theory.

In the case of the Gaussian disorder, we use the results of \cite{stanford} where the required behavior of the spectral density at $E\to -\infty$ is identified with the contribution of small loops in particular region of $(\beta,p)$ plane. It turns out that in such a case, the strength of the disorder is fixed by the boundary value of the dilaton. We suggest the qualitative explanation of that relation in terms of the $T\bar{T}$ deformed boundary quantum mechanics. The Gaussian disorder reminds such deformation and the deformation parameter fixes the radial cut-off in 2D geometry. On the other hand there is the relation between the dilaton boundary value and the radial cut-off which fits with the correct LT dependence on disorder  for $D=1$. Note that the curvature in our derivation does not play any role since this regime corresponds to the small loops which do not feel curvature.

The derivation of the Lifshitz tail for the Poissonian disorder at $E\to 0$ is more subtle. We consider the ensemble of 2D "stretched" (or "inflated") random walks evading the disc of some radius $R$ in the radial geometry. Regarded setting is equivalent to the "magnetic" random walks in the framing with the radial cutoff. Such a geometry brings the sub-ensemble of trajectories to the large deviation regime with the non-Gaussian fluctuations of Kardar-Parisi-Zhang (KPZ) type controlled by the critical exponent $\gamma = \frac{1}{3}$. The KPZ fluctuations are responsible for emergence of the Lifshitz tail in the Poisson disorder in the dual ensemble at large cut-off radius $R$. Our arguments are based on the solution to the radial 2D diffusion equation for the random walk of length $\beta\equiv L=a N$ with the cut-off at the impenetrable disc of radius $R$ upon the "stretching" condition $L=cR$, where $a$ and $N$ are time step length and the number of steps in the path on the two-dimensional lattice. To make our setting relevant to the one discussed in \cite{stanford}, we have  studied the fluctuations of an "inflated" 2D magnetic random ring which is leaned on the impenetrable disc of the radius $R$ and demonstrate the emergence of the KPZ fluctuations.

We will use the picture of the trajectories wrapped around a large cut-off disc for simplicity but the same KPZ scaling takes place for an ensemble of stretched paths staying \emph{inside} a large impenetrable circle close to its inner boundary. Such a picture is relevant for a holography and two settings (paths outside and inside the circle) are related by inversion. In both settings the localization of trajectories in vicinity of the boundary is due to specific value of the magnetic field when the Larmour radius is close to the cut-off radius which ensures stretching of paths compared to the disc radius.

The appearance of the KPZ scaling in the polymer setting is well known (see, for example the review  \cite{halpin1}). The KPZ equation for the growth problem gets mapped onto the directed polymer problem via the Cole-Hopf transform. The height function $h(t,x)$ in the growth problem transforms into the free energy $F=\ln Z$ of the directed polymer, while the time $t$ should be identified with the length of the polymer, $\beta$. The late time  behavior for the directed polymer in the random environment $F=a\beta + c\beta^{1/3}$ is known for a while \cite{dotsenko1, dotsenko2,ledoussal}. In our case similar behavior for the free energy is obtained not from the interaction with the disorder but from a combination of two factors: external magnetic field and the hard boundary. Whether such a "disorder-free" KPZ-type behavior can be seen for (2+1)D disorder polymer problem like studied in \cite{halpin2}, still is an open question.

Recently it was questioned if the holographic duality between the quantum gravity in the bulk
and some boundary theory occurs only when some statistical ensemble of disordered theories at the boundary is considered. For instance, JT gravity is shown to be dual to the very peculiar random matrix model \cite{cotler,sss,sw,mertens1} providing the effective integration over the moduli space of the 2D geometry. The very mapping of the spectral density of the disorder-averaged boundary statistical system onto the spectral density of low-dimensional quantum gravity seems to be one of the most straightforward and transparent issues in the holographic duality. In general, the identification of the quantum gravity as a kind of the random ensemble, is not a simple issue, however there are clear indications that such an identification is indeed required to restore the unitarity of the theory and to reproduce the Page curve in the problem of black hole evaporation \cite{shenker, maldacenaworm,marolf}.

In our study we will emphasize that indeed the ensemble averaging is the key element for the dual representation of the LT even we do not specify the boundary theory. To fit with the logic of boundary matrix model representation of JT gravity in \cite{sss, sw} we point out that LT in $D=1$ can be derived \cite{khorun} from the ensemble of Erd\H{o}sh-R\'{e}nyi exponential random graphs which, to some extent, can be considered as the peculiar discrete version of GOE matrix ensemble. The LT emerges from the fact that in the vicinity of the spectral edge the very special structure of adjacency matrices in form of linear chains  provides the major contribution. That contrasts with the Tracy-Widom behavior near the spectral edge for generic matrix ensembles, where the reduction to the linear structures does not occur.

The paper is organized as follows. In Section II we overview the standard derivations of Lifshitz tails for Gaussian and Poissonian disorders. In Section III we remind the description of JT gravity with the cut-off in terms of the ensemble of magnetic random walks in 2d geometry. Section IV is devoted to the derivation of the specific regimes in JT gravity with the cut-off
for Gaussian and Poissonian disorder   which holographically correspond to the LT near the spectral edge. In Discussion we summarize obtained results and focus on open questions. In Appendices we provide the details of derivations of LT for different models.

\section{Ways to Lifshitz tails}

\subsection{Lifshitz tail in Gaussian disorder: Instanton approach}

The tail of the one-particle spectral density, $\rho(E)$ at large absolute energies for Gaussian disorder $E$, is exponentially decaying, which implies its interpretation via the saddle point for some effective action. Such a generic argument is indeed valid, and the corresponding effective action has been derived in the works \cite{cardy,halperin,luttinger,Yaida,langer} which up to particular details, all rely on the instanton approach. It seems instructive to recall the main features of the instanton computations of the density of states of a $D$-dimensional disordered system and apply them to the system with the $\delta$-correlated random potential.

The logic of the instanton approach is as follows \cite{langer}. We start with the relation between the density of states and the averaged Green function $G_E(x,x'|V)$ at energy $E$
\beq
\rho(E)= \lim_{\epsilon\rightarrow 0} \frac{1}{\pi}\la G_{E+i \epsilon}(x,x|V) \ra
\eeq
where the brackets correspond to the Gaussian averaging over potentials $V(x)$. At the next step
we take the representation of the Green function
\beq
G_{E}(x,x'|V) = \sum_n\frac{\Psi_n(x|V)\Psi^{*}_n(x'|V)}{E_n(V)-E}
\eeq
and assume that there is the saddle potential $\bar{V}$ in the Gaussian integration over V which supports the isolated ground state $\Psi_0$ with energy $E_0$ close to E. Under this assumption the nonlinear equation for $\Psi_0$ can be derived from the initial equation for the Green function whose normalized solution provides the correct expression for the LT.

To be more specific, consider the system described by the Hamiltonian
\be
H = \nabla_{\bf x}^2 + V({\bf x}), \qquad \la V({\bf x}) V({\bf x}')\ra = 2\sigma \delta({\bf x}-{\bf x}')
\label{eq:h}
\ee
where $\nabla_{\bf x}^2$ denotes the $D$-dimensional Laplacian and $\sigma$ controls the Gaussian disorder strength. The effective action reads
\beq
S_{eff} = \frac{1}{\sigma} \int d^Dx V^2(x) - \frac{1}{\sigma} \int d^Dx \lambda\left(E+\nabla_{\bf x}^2-V(x)\right)\Psi + \mu \int d^Dx\left(\Psi^2 -1\right)
\eeq
Upon the extremization of the effective action, one gets
$$
\lambda(x)=C\Psi (x), \quad V(x)=-\lambda(x)\Psi(x)
$$
and for the spherically symmetric case the ansatz for the normalized ground state $\Psi_0(x)$ is
\beq
\Psi_0(x)= \sqrt{\frac{-e}{C}}f\left(\sqrt{-E}(x-x_0)\right)
\eeq
The remaining constraint $\Psi^2 =1$ yields the nonlinear $D$-dimensional Schr\"{o}dinger equation for the normalized ground state wave function $f(r)$
\be
\frac{d^2 f(r)}{d r^2}+\frac{D-1}{r} \frac{d f(r)}{d r} - f(r) + f^3(r) = 0
\label{eq:f}
\ee
with the boundary condition $d_r f = 0$ at $r=0$ and convergence at $r\to \infty$, i.e. $f(r\to\infty) \to 0$. Note that the sign in front of the nonlinear term \eq{eq:f} describing the effective self-interaction, corresponds to the attraction. The equation \eq{eq:f} defines the partition function of the ensemble of self-interacting random walks described by the effective $\phi^4$ scalar field theory with attractive interactions. The final asymptotic expression for the $D$-dimensional spectral density, $\rho(E)$, evaluated via the instanton approach near the band edge, reads
\be
\rho(E) \propto E^{\frac{D(5-D)}{4}} \exp\left(-\frac{\rm const}{\sigma}E^{\frac{4-D}{2}}\right)
\ee
Thus, for $D=1$ one gets
\be
\rho(E) \propto E \exp\left(-\frac{\rm const}{\sigma}|E|^{\frac{3}{2}}\right)
\label{eq:3/2}
\ee

The saddle point solution to the sigma model for the Gaussian disorder selects self-consistently the potential $\bar{V}(x)$, therefore in the leading approximation one could say that a probe particle in a vicinity of the spectral edge propagates in the self-consistent external potential. Let us emphasize the important point which will be used later. The instanton solution implies that the probe particle in the LT regime is strongly localized at some radius in $D$ space coordinates and therefore is localized at point in $D=1$. Remark that the preexponential factor in the LT can be correctly reproduced in the instanton approach as well from the quadratic fluctuations around the instanton solution.

\subsection{Lifshitz tail in Poissonian disorder: Optimal fluctuation and survival probability}

The Lifshitz tail in the Poissonian disorder admits two related descriptions which clarify the physical origin of the phenomenon. The first, mostly qualitative derivation, goes as follows \cite{rodgers,neuwen1}. Consider the linear chain consisting of some number of elementary bits connected via random bonds of some nature \cite{rodgers}. Let $N$ be the total number of linked bits and the attachment of the additional bit costs $e^{-p}$. We are interested in the low energy tail of the spectral density of a string of $N$ connected bits, which can be written as
\beq
\rho(E) \propto \sum_{N}^{\infty} (pe^{-p})^N \delta\left(E-\frac{1}{pN^2}\right)
\label{e:bit}
\eeq
At each bond joining neighboring vertices, the non-relativistic particle propagates freely which explains the argument in the delta function in \eq{e:bit}. Equation \eq{e:bit} with the free boundary conditions immediately provides the Lifshitz tail in $D=1$ at $E\to 0$.

These simple arguments show that we deal with the behavior of the composite trajectory of a particle fluctuating in random "cages" with the Poissonian distribution, or with the long string built of randomly connected elementary bits. In Appendix \ref{app2} we present two disorder-free models where gluing bits is made precise, and the curvature replaces the effect of the disorder. The corresponding picture has a lot in common with the formation of the long string from the short ones or its decay near the horizon where the string is almost tensionless. Such processes have a similarity with the Hagedorn transition \cite{susskind,mertens}.

Another derivation of the Lifshitz tail in the Poissonian disorder deals with the "survival probability", and can be considered as the kind of an inverse Laplace transform of \eq{e:bit}. The investigation of the survival probability in an ensemble of random walks in a Poissonian field of random traps has been proposed in the pioneering works of Balagurov and Vaks \cite{balagurov}, was rigorously studied by Donsker and Varadhan \cite{donsker}, and later appeared in the literature in numerous incarnations (see for example \cite{mova, seif}. The trapping problem has an exact solution in $D=1$. Since the scaling consideration seems more illuminating for our purposes, we briefly recall the derivation of the survival probability via the method of optimal fluctuations.

Consider the one-dimensional $N$-step biased random walk on the interval of length $h$. Suppose that the probability $Q(h)$ to find an interval of length $h$ is Poissonian, which up to logarithmic corrections can be approximated by the exponential distribution, $Q(h)\sim e^{-\kappa h}$, where $\kappa$ ($\kappa=\mathrm{const}>0$) controls the distribution width (the disorder strength). We are interested in the typical "survival" probability, $W(N)$, for a biased random walk to hit the boundary for the first time at the very last step, $N$, in the ensemble of intervals distributed with $Q(h)$. The non-equilibrium free energy, $F(N,h)$, of the system consists of two parts: $F_f$ ("fluctuational") due to the free random walk within the given interval, $h$, and $F_d$ ("disordered") due to the random distribution of $h$. For $F_f$ we have the following estimate
\be
F_f(N,h) = N\lambda_{min}(h)+ N u
\label{eq:f(f)}
\ee
where $\lambda_{min} = \frac{2\pi a^2}{h^2}$ is the minimal eigenvalue of the Laplacian on the interval $L$ with the Dirichlet boundary conditions, $u$ is the dimensionless bias (the drift field), and $F_d(h)=-\ln Q(h)$. Minimizing the non-equilibrium free energy
\be
F(N,h) = F_f(N,h) + F_d(h) = \frac{2\pi Na^2}{h^2} + N u + \frac{\kappa h}{a}
\label{eq:free}
\ee
with respect to $h$, we get $h_{opt} = \left(\frac{4\pi a}{\kappa}\right)^{1/3} N^{1/3}$, where $a$ is the step length of the random walk. Thus, the survival probability $W(N) = \exp(-F(N,h_{opt}))$ reads
\be
W(N) \propto \exp\left(-Nu - \gamma N^{1/3}\right)
\label{eq:surv}
\ee
where $\gamma=\frac{3}{2}(4\pi\kappa^2)^{1/3}$.

Comparing \eq{eq:surv} to the 1D case \eq{eq:n(L)}, we conclude that at $D=1$
\begin{multline}
\rho(E) \propto \frac{1}{2\pi i}\int\limits_{\eps-i\infty}^{\eps+i\infty} W(N)\, e^{NE} dN \propto \frac{1}{2\pi i}\int\limits_{\eps-i\infty}^{\eps+i\infty} \exp\left(-Nu - \gamma N^{1/3}\right) \, e^{NE} dN \propto \\ \exp\left(-\frac{\gamma^{3/2}}{\sqrt{E-u}}\right)
\label{e:03a}
\end{multline}
The constant $c_{D=1}$ in \eq{eq:n(e)} can be identified with $\gamma^{3/2}$ in \eq{e:03a}.

The outlined "optimal fluctuation" derivation of the Lifshitz tail \eq{e:03a} is a variant of an indirect instanton approach for treating the 1D particle localization in a quenched random potential with a Poissonian disorder.

\subsection{Lifshitz tail in Poisson disorder: Instanton approach}

The direct instanton approach can be developed for the Poissonian disorder as well, however
the construction of the effective potential in the corresponding sigma model is more elaborated than that for the Gaussian disorder. Following \cite{luttinger} assume that the interaction between a non-relativistic particle and an impurity is repulsive and short-range (i.e. the interaction radius is shorter than any other lengths in the model). Using the replica trick, it is possible
to derive the effective action, $S(\phi)$, and the corresponding spectral density, $\rho(E)$,
\beq
\rho(E) = \lim_{n\to 0} \mathrm{Im} \frac{\rho}{\pi E n}\int d^n\phi\, \left(\phi \sqrt{E}x\right)^2 \exp\left(-\rho E^{-3/2}S(\phi)\right)
\eeq
where
\beq
S(\phi)=\int d^Dx \left(1 - e^{-\chi \phi^2} - \phi(1 +i\epsilon + \nabla_r^2)\phi \right)
\label{replica}
\eeq
In (\ref{replica}) the field $\phi$ is the vector in the replica space, $A$ is the disorder strength, $\chi=\frac{\pi \bar{C} A}{E}$, and $\bar{C}$ is the density of impurities. Evaluating the
action \eq{replica} at the instanton solution, one arrives at the following expansion for the $D$-dimensional spectral density
\beq
\rho(E)\propto \exp\left(-\left(c_1 E^{-\frac{1}{2}}+c_2E^{-\frac{3}{2}}+...+c_DE^{-\frac{D}{2}}
\right)\right)\Big|_{E\to 0} \to \exp\left(-c_D E^{-\frac{D}{2}}\right)
\eeq

The effective action, $S(\phi)$, can be generalized by introducing an additional parameter, $p$, which defines the probability for a lattice site to have a trap \cite{lubensky, renn}. The potential term in the effective action for the replica field upon such a modification reads
\beq
U(\phi) =\phi^2 - \ln\left(p +(1-p) e^{-\chi \phi^2}\right)
\eeq
which in the limit $p\to 1$ reduces to the potential term in the action (\ref{replica}).

\subsection{Lifshitz tail in spectral density of Erd\H{o}sh-R\'{e}nyi random networks}

To identify the universality class of the disordered system it is useful to map it to the particular matrix model at large $N$. It is eligible to ask a question which is the simplest matrix model (MM) pattern involving the LT at the spectral edge. To proceed, it is suitable to step away from the matrix model paradigm and consider an ensemble of exponential random graphs. Namely, the following replacement is implied:
\beq
Z(V)= \int dM e^{\Tr V(M)} \quad \to \quad Z(V) = \sum_{\rm graphs} e^{\Tr V(M)}
\label{MM}
\eeq
where in the MM framework the matrix $M$ belongs to some known matrix ensemble ($GOE$ for example), while at the random network side, $M$ is the adjacency matrix of the graph. The potential $V$ can be arbitrary and in the network case it is well known that $\Tr M^k \propto \#\{\mbox{loops of size $k$}\}$. Hence, different times in the matrix models correspond to the different chemical
potentials for the number of loops.

The simplest case of the Gaussian matrix model with a potential $V(M)=M^2$ corresponds to the Erd\H{o}sh-R\'{e}nyi (ER) network with the chemical potential the number of links (called sometimes "2-cycles"). We are interested in the spectral density of the graph Laplacian defined as $L= D-M$, where $M$ is the adjacency matrix and $D=\mathrm{diag}(d_1,\dots d_N)$ is the matrix of vertex degrees. The spectrum of the graph Laplacian, $L$, is positive and the degeneracy of the minimal eigenvalue, $\lambda=0$, corresponds to the number of disjoint graph components. The Lifshitz tail emerges at $\lambda\to 0$. It has been shown rigorously in \cite{khorun} that in the vicinity of the spectral edge the main contribution to the spectral density originates from sparse graphs and the majority of them are weakly branched, i.e have presumably linear structure. In more details the topology of subgraphs for sparse adjacency matrices has been discussed in \cite{krapiv}. At the edge $\lambda=0$, the spectral density has 1D Lifshitz tail -- exactly as in the (1+1)D disordered system, despite that the Erd\H{o}sh-R\'{e}nyi random network definitely is not a one-dimensional object. However, as it has been shown in \cite{kovaleva}, even the ensembles of tree-like graphs have the Lifshitz tail in the spectral density. For ensemble of sparse graphs the emergence of the Lifshitz tail has been also discussed in \cite{krapiv,polov} in connection with the ultrametric structure of the Riemann "raindrop" function, related to the modular Dedekind $\eta$-function. The strength of the disorder, $c_1$, in the spectral density $\rho(\lambda) \propto \exp(\frac{c_1}{\sqrt{\lambda}})$, at the spectral edge, $\lambda\to 0$, can be identified with the link formation probability in the Erd\H{o}sh-R\'{e}nyi model.

The existence of LT in the exponential random network is very sensitive to clustering. It is known in the network context that isolated soft eigenvalues of the graph Laplacian, $\hat{L}$, correspond to clusters which are the precise counterparts of the eigenvalue instantons in matrix models. In presence of several clusters, instantons interact and form the non-perturbative soft band in the spectral density, which modifies the Lifshitz tail. That behavior has been seen in \cite{eigen} when by imposing the non-singlet constraint in the ER model and assigning the chemical potential for the 3-cycles, the formation the second non-perturbative band at $\lambda\to 0$ was induced above some critical value of the chemical potential.

\section{JT gravity at finite cutoff: Lifshitz tail in Gaussian disorder}

In this Section we argue that emergence of LT in $D=1$ system with the Gaussian disorder can be reproduced holographically in the flat JT geometry in the particular region of $(\beta,p)$ plane. The boundary value of the dilaton which yields the effective pressure in the polymer representation, can be associated with the disorder strength in the boundary QM. The small loops are responsible for the LT in this case. The Lagrangian of the JT gravity \cite{j,t} involves the metric and the dilaton field, $\phi$
\beq
L_{JT}= -S_0\, \chi(M) - \int_M \phi\left(R+\frac{2}{l^2}\right) - 2\int_{\partial M}\phi\, K
\eeq
where $\chi(M)$ is the Euler characteristic of the manifold $M$, $l$ is the $AdS_2$ radius and the
last term is the boundary Gibbons-Hawking term. We are interested in the partition function of the JT gravity in the disc with the fixed boundary length, $L$, and fixed value of the boundary dilaton, $\phi_b = p\,l^2$
\beq
Z(L,p)= \int_{L,p} Dg_{\mu\nu} D\phi e^{-S_{JT}}
\label{eq:z}
\eeq
Variation of the Lagrangian with respect to the dilaton yields
\beq
R=-\frac{2}{l^2}
\eeq
which defines the two-dimensional hyperbolic disc and the JT action reads
\beq
S_{JT}= -2\pi p\,l^2 - p\,A
\eeq
where $A$ is the disc area. Therefore the path integral \eq{eq:z} gets reduced to the path integral over loops $\Gamma$ weighted with the enclosed area in $AdS_2$:
\beq
Z(L,p)=\int \frac{D\{{\rm paths}~\Gamma\}}{SL(2,R)}\, e^{p\,A +2\pi p\, l^2}
\label{path}
\eeq
which runs over the space factorized by the action of the symmetry group $SL(2,R)$. Splitting the loop into $N$ straight segments of length $a$ each, we can express the renormalized length $\beta$ in terms of $N$, $a$ and $l$:
\beq
\beta= \frac{Na^2}{2l}
\eeq

Remarkably, it was found in \cite{kitaev,yang} that the partition function $Z_{JT}(\beta)$ has an appealing realization as the partition function of a particle propagating in Euclidean time in the hyperbolic plane $H_2$ in the effective external transverse magnetic field
\beq
Z_{part}(\beta)= \sum_{\rm paths~\Gamma} e^{p\, A}
\eeq
The partition function $Z_{part}(\beta)$ can be evaluated exactly due to the high symmetry of the problem. The explicit relation between the $Z_{JT}$ and $Z_{part}$ is as follows:
\beq
Z_{JT}(\beta,p)= Z_{part}(\beta,p)
\eeq
The partition function $Z_{JT}(\beta,p)$ can be considered as the Laplace transform of the spectral density
\beq
Z_{JT}(\beta)=\int dE \rho(E) e^{-\beta E}
\eeq.

Various regimes in the $(\beta,p)$ plane  amount to the different forms of  the spectral density, $\rho(E)$, which have been identified in \cite{stanford}. When typical loops are relatively small ($\beta \lesssim 1$), the curvature can be neglected, and the problem gets reduced to the evaluation of the partition function of self-interacting polymers in flat space with an external pressure. This partition function has been computed in \cite{richards} and can be expressed in terms of the logarithmic derivative of the Airy function. The corresponding Laplace-transformed partition function (i.e. the spectral density) reads
\beq
\rho(E)\propto \frac{p^{1/3}}{{\rm Ai}^2(-Ep^{-2/3}) + {\rm Bi}^2(-Ep^{-2/3})}
\eeq
The limit of the spectral density at $E\to -\infty$ reads as
\beq
\rho(E) \propto \exp\left(-\frac{E^{3/2}}{p}\right)
\eeq
which coincides with the Lifshitz tail in $D=1$ for the Gaussian disorder upon the identification $\sigma^{-1} \Leftrightarrow p$, where $\sigma^{-1}$ is a disorder strength in the boundary disordered system.

Is the identification $\sigma^{-1} \Leftrightarrow p$  between the disorder strength in QM and the pressure in the dual gravity self-consistent? To answer this question qualitatively, let us recall the relation between the JT gravity at finite cutoff and the $T\bar{T}$ deformation \cite{zamolodchikov} of the boundary (0+1) theory \cite{gross,verlinde}. Consider the deformation of the Hamiltonian in QM by the term $\lambda T^2$ where $T$ is the stress tensor. For static configurations, $T$ can be identified with the potential $V(x)$, hence the term $\sigma^{-1} V^2$ corresponding to the Gaussian disorder has exactly a requested form: $\lambda=\sigma^{-1}$. Some reservations are in order. Note that in the disorder problem we integrate over the potentials $V(x)$ with the Gaussian measure while there is no such immediate integration over the stress tensor in the $T\bar{T}$ deformation problem. Secondly, the identification of the potential and the stress tensor is based on the instanton solution strongly localized in $D=1$ however the effects of fluctuations around the instanton solution certainly involve the kinetic term.

Assuming the relation $\lambda=\sigma^{-1}$ we can use the relation between $\lambda$ and the cut-off scale $r_c$ which was found in \cite{gross}
\beq
\lambda =\frac{2\pi G}{\phi_b r_c^2}
\label{relation}
\eeq
where $r_c$ is the cut-off radius with the Dirichlet boundary conditions and $\phi_b$ is the renormalized boundary value of dilaton. Note that we use  (\ref{relation}) since it takes into account the embedding into 3D which is necessary in our problem of LT in $D=1$. Indeed in spite of consideration in $(t_E,r)$ space we have take into account in the matching condition one extra space dimension. Emphasize once again that the problem essentially gets reduced to $(t_E,r)$ plane due to the localization at the instanton solution in the effective sigma model.

Assembling these qualitative arguments we argue that the identification of the boundary disorder with the bulk magnetic field, $\sigma^{-1}\propto \phi_b\propto p$, is self-consistent if the finite cutoff, $r_c$, is assumed. Note that the identification boundary value of the dilaton with the strength of disorder in the boundary theory is somewhat similar to the SYK case in the Schwartzian low-energy limit.

\section{JT gravity at radial cutoff: Lifshitz tail in Poissonian disorder via KPZ fluctuations}

Here we extend our analysis to the boundary quantum mechanics with the Poisson disorder for uncorrelated defects. Once again we focus at the $D=1$ case. Proceeding similarly to the consideration of the Gaussian disorder, we introduce the cut-off in the radial coordinate, however here the cutoff is less justified than in the Gaussian case since there is no argument based on the $T\bar{T}$ deformation. We argue that the LT can be reproduced if one considers a very special regime in the polymer representation of the 2D dilaton gravity. First, note that small energies (small eigenvalues in the spectral description) correspond to large $\beta$ hence we could expect that large lengths of loops in 2D dominate.

Thus, we have the radial cutoff, $r_c=R$, the typical loop lengths, $\beta$, and the value of the magnetic field weighting the area, $A$, enclosed by the loop. How all these variables are related?
We shall focus at the special limit when the value of the magnetic field which provides the Larmour radius of the charged particle is chosen in such way that Larmour radius is of order of $R$. Therefore, in that case the contributions to the partition function (written as a path integral) coming from of the trajectories with fixed length, $\beta$ (but with different areas), obey the relations:
\be
\begin{cases}
\beta=cR & \mbox{where $c \sim 1$} \\
p \propto R^{-1}
\end{cases}
\label{cond}
\ee
In addition to \eq{cond}, it is assumed that $R \gg 1$.

As it follows from \eq{eq:n(L)} (see also \eq{e:03a}), to get the Lifshitz tail in the spectral density upon the Laplace transform, the partition function, $Z(\beta)$, should have at $\beta \gg 1$ the asymptotic behavior
\beq
Z(\beta)\propto \exp\left(-\mathrm{const}\, \beta^{1/3}\right)
\label{13beta}
\eeq
The question to be answered behind this formal relation is as follows: could we see a Griffiths-like behavior \eq{13beta} of the partition function in the regime which we have described above? We provide some arguments, using recent results derived in \cite{nechaev1, nechaev2, gnv} concerning the KPZ scaling of fluctuations in the restricted random walks on the plane. To justify the Griffiths-like behavior we first regard the model A of restricted random walk of fixed length, $L$, where the relation between the length, $L$, and the cutoff scale, $R$, is imposed by hands.

\begin{figure}[ht]
\includegraphics[width=16cm]{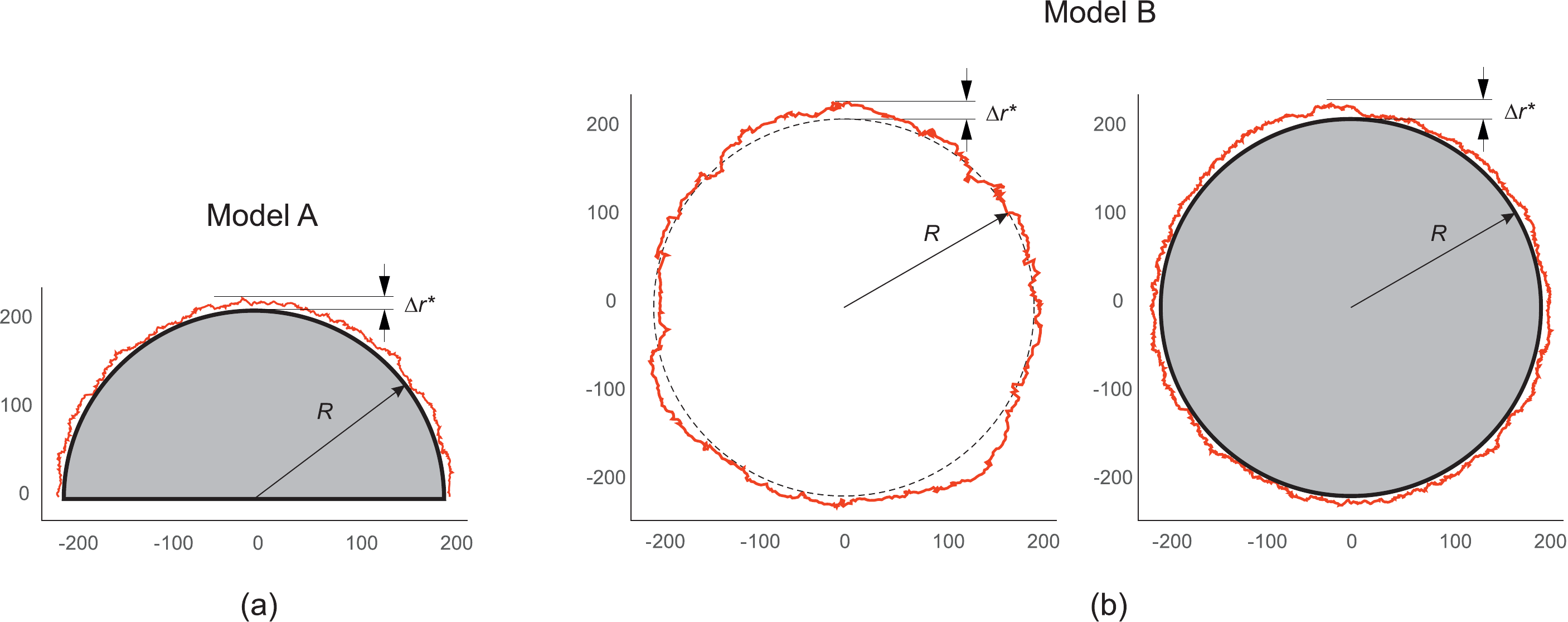}
\caption{Models A and B of stretched paths above the impenetrable circular domain: a) In the model A stretching is ensured by imposing by hands condition $L=cR$ on the relation between the disc radius and the path length; b) In the model B the random loop is strongly inflated with a large algebraic area, as it is shown by dotted line in the left panel of figure (b), and then inside the inflated loop the impenetrable disc of the radius $R$ is inserted such that the path is leaning on a disc boundary as it is shown in the right panel of figure (b).}
\label{fig:AB}
\end{figure}

The model A, shown in \fig{fig:AB}a is defined as follows. Consider the stretched two-dimensional random walk of length $\beta\equiv L=a N$ (without the magnetic field) evading the impenetrable semicircle of radius $R$. Stretching of the walk is ensured by imposing the linear relation, $L = cR$ by hands, where $c$ is some positive constant. We consider the limit $N\gg 1$ (respecting the condition $L=cR$) and show that there is an interval of values of $c$ for which the Brownian trajectories are localized near the disc boundary within the depletion layer $[R, R+ \mathrm{const} R^{1/3}]$. At the value $c_{tr}$ the typical width of fluctuations experiences the transition: at $c<c_{tr}$ the fluctuations are controlled by the KPZ critical exponent, $\gamma=\frac{1}{3}$, while at $c>c_{tr}$ the fluctuations are Gaussian with $\gamma=\frac{1}{2}$. The model A supports the conjecture that the desired KPZ scaling can be derived from the bunch of trajectories localized near the disc boundary.

To proceed further, define the model B (see \fig{fig:AB}b) and consider the random loop of length $L$. Here the condition $L=cR$ is not "imposed by hands", but is selected dynamically. To this aim we switch on the external magnetic field and adjust is strength such that it ensures the "inflation" of regarded loops. Adjusting the area, $A$, enclosed by the "inflated" loop of length $L$, with the typical gyration radius of the loop, $R_g$, we can ensure the stretching condition, $L \propto R_g$. That is, we consider the ensemble of two-dimensional $N$-step Brownian loops in the magnetic field with a strongly inflated algebraic area enclosed by the loop. Then, we insert inside the inflated loop the impenetrable disc of radius $R=cL$ and adjust $c$ such that the Brownian loop is "leaning" on the disc boundary. In such a setting, we see again the regime with KPZ fluctuations. We argue that the combination of an effective pressure and the cutoff due to the impenetrable boundary of the disc, ensures the emergence of the KPZ fluctuations. Computing the typical span of the path's fluctuations above the disc, we are able to estimate the typical free energy of the ensemble of trajectories, which permits us to derive the LT in the Poissonian disorder. Remind that in the holographic setting magnetic field corresponds to the boundary value of the dilaton.

In \fig{fig:AB}a,b we show typical samples of Brownian paths obtained in computer simulations for models A and B. The average span of path's fluctuations above the boundary is denoted by $\Delta r^*$. In the model A we consider a trajectory stretched above a half of a disc. Such a setting being eligible due to the symmetry of the problem, permits essentially speed up the computational time.

The details of derivation of the Griffiths-type contribution to the partition functions for models A and B are presented in the Appendix \ref{app2}, where the problem is reduced to the solution the two-dimensional diffusion equation in stretched regime subject the boundary condition at the cutoff disc. It should be emphasized that for the model B the presence of both factors: (i) the magnetic field ensuring stretching, and (ii) the boundary which "supports" the stretched trajectories, is crucial for emergence of anomalous fluctuations. The absence of at least one of these conditions, returns the span of fluctuations, $\Delta r^*$, to the Gaussian regime. The evaluation of the partition function with the fixed length in the described regime yields
\beq
Z(N) \propto e^{-(a/R)^{2/3}N^{1/3}}
\eeq
where $R$ is the curvature radius of a disc boundary, $a$ and $N$ are the length and the number of steps in the path stretched around the disc (recall that $a N \equiv \beta$). The spectral density, $\rho(E)$, is derived by the saddle point evaluation of the Laplace transform of $Z(N)$. To check the validity of the saddle point we have to verify that indeed the contribution from $\beta \sim R$ matters. Looking at the typical energies consistent with these conditions we immediately see that $E\rightarrow 0$ as expected for the LT.

Let us emphasize that from the holographic viewpoint one has to consider the trajectories lying in the skin layer inside the disc with cutoff at the disc boundary (contrary to the trajectories located outside the disc as it is shown in figures \fig{fig:AB}). However the ensembles of stretched trajectories close to the disc boundary \emph{from both sides} manifest the KPZ scaling and are related by the simple inversion argument.

\section{Discussion}

In this work we probe the holographic duality for a non-relativistic quantum mechanics focusing at the Lifshitz tail in the one-particle spectral density  in the disordered environment in $D=1$. Since the LT phenomenon is quite universal, we have asked (without specifying the details of a boundary quantum mechanics) whether the LT in the spectral density can be  extracted holographically from the $(t_E,r)$ part of the whole bulk geometry. We have conjectured that the LT in $D=1$ both for Gaussian and Poisson disorder can be determined holographically by taking some limits of 2D quantum dilaton gravity with cut-off. We make some modest step towards evidences for this conjecture using the polymer picture for partition function of 2D dilaton gravity developed in \cite{kitaev,yang, stanford}. Our main object is the ensemble of fluctuating inflated polymer loops, or charged particle trajectories in the 2D geometry in presence of an external transverse magnetic field and radial cut-off.

It turns out that the analysis for Gaussian and Poissonian disorders is a bit different, however in both cases we have argued that a presence of a radial cut-off is crucial.
\begin{enumerate}
\item[(i)] For the Gaussian disorder, the emergence of the cut-off can be argued by analogy with the $T\bar{T}$ deformation in the boundary quantum mechanics which is known to be holographically dual to a theory with a cut-off. In that case the bunch of \emph{small} loops provides the dominant contribution to the LT, and the disorder strength is related with the dilaton boundary value. The underlying 2D geometry is essentially flat.
\item[(ii)] For the Poissonian disorder, we have introduced the radial cut-off as well, and in that case the dominant contribution to the partition function comes from the ensemble of \emph{large} inflated Brownian loops fluctuating in the vicinity of the large impenetrable cut-off disc in the 2D radial geometry. The typical length of the trajectory in the effective magnetic field is comparable to the length of the cut-off disc boundary. That is why the dominant contribution to the partition function comes from the trajectories staying nearby the disc boundary. The key property, responsible for the emergence of the LT, is a very peculiar property of such class of paths: they share the KPZ scaling, which is Laplace dual to the LT in the spectral density. We have  discussed the $D=1$ case, where the physical origin of the LT for Poissonian disorder is connected with the linear topology of subgraphs near the spectral edge.
\end{enumerate}

The reader could raise the legitimate question how the space disorder from defects distributed in $D=1$ manifests itself in the dual holographic theory in $(t_E,r)$ space with  boundary $t_E$ coordinate where there is no space direction at all. We have conjectured that qualitatively the space disorder gets mapped onto random distribution of "renewal times" at the Euclidean time coordinate running along the disc boundary. Let us remind that minimization of \eq{eq:free} provides the typical size of a random "cavity", $h_{opt} = \left(\frac{4\pi a }{\kappa}\right)^{1/3} N^{1/3}$, within which the random trajectory is localized. The typical "renewal time", $t_{opt}$, is the time which random walk spends within the optimal cavity, $h_{opt}$. Since the path within $h_{opt}$ is a random walk, one can estimate $t_{opt}$ as follows
$$
t_{opt} \sim \frac{h^2_{opt}}{a^2}\sim \kappa^{2/3} N^{2/3}
$$
Thus, the full time, $N$, splits into a sequence of $\frac{N}{t_{opt}}\sim \frac{N^{2/3}} {\kappa^{2/3}}$ independent renewal times, $t_{opt}$. These renewal times are consistent with the "polymer inflation" interpretation ensured by the magnetic field acting on a charged particle in 2D near the disc boundary -- see Appendix \ref{app2}.

Another point concerns the generalization of our 1D picture for the LT to $D$-dimensional space. In that case the effective reduction to $(t_E,r)$ bulk coordinates does not work and higher-dimensional bulk involving space coordinates has to be considered. We hope to return to this point in a separate study. To this aim it is necessary to clarify the role of the effective sigma
model and the corresponding instanton solution with finite action in the holographic setting. Certainly the underlying reason for two different representations of the LT is due to two representations of the partition function: via the trace over the Hilbert space and via the path integral over trajectories in the phase space. However it is not clear how this general argument works exactly in the holographic representation of LT.

The representation of the 2D dilaton gravity partition function via the particle in the magnetic field raises one more question. The magnetic field in the Euclidean time can be considered as the electric field upon the Wick rotation. Closed trajectories in the Euclidean time can be considered as the bounce configurations for the Schwinger pair creation in the electric field. Since we look at the trajectory of the particle, the relevant question concerns the induced Schwinger pair creation when the particle trajectory can be decorated by small circles representing the bounces. In principle, such a non-perturbative process could provide the source of disorder at the dual gravity side and contribute to the LT. It would be interesting to investigate this possibility.

In the case of the Poisson disorder the derivation of LT effectively reduces to the summation over the ensemble of trajectories between two boundaries in the radial geometry: the lower boundary, $r_-=R$ is the radius of the impenetrable disc, $R$, on which the inflated loop is leaned, while the upper boundary, $r_+$, is the radius of the maximally inflated loop of $N$ steps, $r_+=\frac{Na}{2\pi}$. To some extend, our consideration is the first quantized representation for the scalar field with two boundary conditions which provides a source for thoughts about the connection between the Lifshitz tail and the Berezinsky-Kosterlitz-Thouless (BKT) critical behavior. The dependence of the order parameter on the temperature near the BKT transition in 2D behaves identically to the LT in (1+1)D theory. The treatment of the BKT transition in the holographic (renormalization group) setting leads to the solution to the Schr\"{o}dinger-like equation for the scalar field in the radial $AdS_2$ framing in the conformal $r^{-2}$ potential acting between the UV ($r_-$) and IR ($r_+$) boundaries \cite{mezei,son}. Such a setting is very similar to our treatment of inflated loops above the impenetrable disc considered in Appendix \ref{app2}.

The model with the Poisson disorder involves the specific linear structure formed from the elementary bits near the boundary of the restricted region. This resembles the formation of the long strings from the short ones near the black hole horizon \cite{susskind,mertens}. In the black hole context this region was related to the stretched horizon. Our discussion suggests that the KPZ
scaling could be relevant for this phenomenon. It would be interesting to pursue this analogue further.

\begin{acknowledgments}

We are grateful to A. Kamenev, A. Milekhin, K. Polovnikov, S. Shlosman, M. Tamm, and A. Vladimirov for numerous discussions and valuable comments. Authors acknowledge the BASIS Foundation for the support within grants (20-1-1-23-1 for AG and 19-1-1-48-1 for SN and AV). The work of AG was supported by grant RFBR-19-02-00214 and by grant 075-015-2020-801 of Ministry of Science and Higher Education of the Russian Federation. The work of AV was supported by the state task for the FRC CP RAS \# FFZE-2019-0016.

\end{acknowledgments}

\begin{appendix}

\section{Lifshitz tail in 1D localization}
\label{app1}

\subsection{Localization in bimodal site disorder}

Let us recall the formulation of the 1D Anderson model. The time-dependent distribution function, $\psi_t(x)$, obeys the discrete Schr\"{o}dinger-type equation in a real time:
\be
\begin{cases}
\psi_{t+1}(x) = \psi_t(x-1) + \psi_t(x+1) + \omega_x \, \psi_t(x) & \mbox{for $x=1,...,L$} \medskip \\ \psi_t(x=0)=\psi_t(x=L+1) = 0 \medskip \\ \psi_{t=0}(x) = \delta_{x,x_0}
\end{cases}
\label{e:16}
\ee
where $\omega_x$ for any $x=1,...,L$ is a random variable with a bimodal distribution characterized by the parameter $p$ ($0<p<1$):
\be
\omega_x = \begin{cases}
m & \mbox{with the probability $p$} \medskip \\
M & \mbox{with the probability $1-p$}
\end{cases}
\label{e:17}
\ee
We suppose that $0<m\ll M$. Passing to the generating function $\Psi(x,s) = \sum_{t=0}^{\infty} \lambda^t \psi_t(x)$, we arrive at the spectral problem, where the "chemical potential", $\ln\lambda$, plays the role of the energy.

The spectral behavior of the Anderson model for the distribution \eq{e:17} up to some minor details is identical to the problem encountered in the study of harmonic chains with binary distribution of random masses. The last problem, which goes back to F. Dyson, has been investigated in \cite{domb} and then thoroughly discussed in \cite{huisen}. They considered a harmonic chain of masses, which can take two values, $m$ and $M>m$ as implied by \eq{e:17}. In the limit $M\to\infty$ and $m=\mathrm{const}$, the system breaks into "uniform harmonic islands" consisting of light masses ($m=1$ by definition) surrounded by two (left and right) infinitely heavy masses, $M$. The probability to have such a harmonic island of $h$ consecutive light masses is $(1-p)^2 p^h$. So, one gets straightforward mapping onto the spectral statistics of ensemble of independent harmonic chains with an exponential distribution of their lengths.

Many results concerning the spectral statistics of ensemble of sequence of random composition of
light and heavy masses were obtained in the works \cite{domb,huisen}. In particular, in
\cite{huisen} the integrated density of states, ${\cal N}(\lambda)$ has been written in the
following form:
\be
{\cal N}(\lambda) = \int_{-\infty}^{\lambda} \rho(\lambda')d\lambda' =
1-\frac{1-p}{p^2}\sum_{n=1}^{\infty} p^{\rm Int\left(\frac{n\pi}{\vartheta}\right)}; \qquad \cos\vartheta=\frac{\sqrt{\lambda+1}}{2} \quad \left(0<\vartheta<\frac{\pi}{2}\right)
\label{e:18}
\ee
For $\lambda\to -1$ one gets ${\cal N}(\lambda)\to p/(1+p)$ which corresponds to the contribution
of states at the bottom of the spectrum, near $\lambda=-1$. At the upper edge of the spectrum (for
$\lambda\to 3^{-}$) one gets ${\cal N}(\lambda)\to 1$ which means that all states are encountered.
Equation (\ref{e:18}) shows that the behavior near $\lambda=3$ (i.e. at $\vartheta=0$) is
dominated by the first term ($n=1$) of the series. Therefore, one has:
\be
{\cal N}(\lambda) \simeq 1-\frac{1-p}{p^2}p^{2\pi/\sqrt{3-\lambda}}
\label{e:19}
\ee
The expression \eq{e:19} signals the appearance of the Lifshitz tail in the density of
states. A more precise analysis shows that the tail \eq{e:19} is modulated by a periodic function
\cite{huisen}. Equation \eq{e:18} displays some interesting features. In particular, the function ${\cal N}(\lambda)$ occurs in the mathematical literature as a generating function of the continued fraction expansion of $\frac{\pi}{\vartheta}$. Let us briefly sketch this connection following \cite{borwein}. Consider the continued fraction expansion:
\be
\frac{\pi}{\vartheta}=\frac{1}{\disp c_0+ \frac{1}{\disp c_1+ \frac{1}{\disp c_2+\ldots}}}
\label{e:20}
\ee
where all $c_n$ are natural integers. Truncating this expansion at some level $n$, one gets a
coprime quotient, $\frac{p_n}{q_n}$ which converges to $\frac{\pi}{\vartheta}$ when $n\to \infty$.
A theorem of Borwein and Borwein \cite{borwein} states that the generating function, $G(z)=\sum_{n=1}^{\infty}z^{\rm Int\left(\frac{n\pi}{\vartheta}\right)}$ of the integer part of $\frac{\pi}{\vartheta}$ is given by the continued fraction expansion
\be
G(z)=\frac{z}{1-z} \frac{1}{\disp A_0+ \frac{1}{\disp A_1+ \frac{1}{\disp A_2+\ldots}}}; \qquad A_n(z)=\frac{z^{-q_n}- z^{-q_{n-2}}} {z^{-q_{n-1}}-1}
\label{e:21}
\ee
and $q_n$ is the denominator of the fraction $\frac{p_n}{q_n}$ approximating the value
$\frac{\pi}{\vartheta}$. In order to connect $G(z)$ with the integrated density, \eq{e:19},
it is sufficient to set $z=f$ and express ${\cal N}(\lambda)$ in terms of $G(z)$.

The spectral statistics of three-diagonal band matrices with the diagonal binary "heavy" disorder is straightforwardly transferred to the random walk survival probability in one-dimensional array of Poissonian traps.

\subsection{Spectral statistics of Schr\"{o}dinger-like operator with random hopping}

\subsubsection{Edge spectral behavior}

Consider an ensemble of random three-diagonal operators $N\times N$ ($N\gg 1$) represented by symmetric random matrices $A_N$ with the bimodal (Bernoulli) distribution of sub-diagonal matrix elements:
\be
A_N = \left(\begin{array}{ccccc}
0 & \eps_1 & 0 & \cdots & 0 \smallskip \\  \eps_1 & 0 & \eps_2 & & \smallskip \\  0 & \eps_2 & 0 & & \smallskip \\ \vdots &  &  &  & \smallskip \\ & & & & \eps_{N-1} \smallskip \\ 0 & & & \eps_{N-1} & 0 \end{array} \right);
\qquad \eps_x=\left\{\begin{array}{ll} 1 & \mbox{with probability $p$} \medskip \\
0 & \mbox{with probability $1-p$} \end{array} \right.
\label{e:04c}
\ee
The matrix $A_N$ at each $\eps_x=0$ splits into regular (gapless) three-diagonal submatrices. The probability to find such a submatrix ("cage") of size $hD$ is $Q(h) = p^h$. Each $h\times h$ cage is a three-diagonal symmetric matrix $A_h$ with all elements $\eps_x$ equal to 1.

The time-dependent distribution function, $\Psi_t=(\psi_t(1),...,\psi_t(x),...,\psi_t(N))^{\top}$, obeys the discrete stochastic equation with a transfer matrix $A_N$:
\be
\Psi_{t+1} = A_N \Psi_t; \qquad  \Psi_{t=0} = (0,..,0, \psi_0(x_0), 0,...,0)^{\top}
\label{e:05c}
\ee
Written in components, $\psi_t(x)$, for $x=1,...,N$, \eq{e:05c} resembles the discrete Schr\"{o}dinger-type equation in a real time:
\be
\begin{cases}
\psi_{t+1}(x) = \eps_{x-1}\, \psi_t(x-1) + \eps_{x}\, \psi_t(x+1)  & \mbox{for $x=1,...,N$} \medskip \\ \psi_t(x=0)=\psi_t(x=N+1) = 0 \qquad (\eps_0=\eps_N=0) \medskip \\ \psi_{t=0}(x) = \delta_{x,x_0}
\end{cases}
\label{e:06c}
\ee
Since all $\eps_x$ ($x=1,...,N$) do not depend on $t$, one can pass from \eq{e:06c} to the generating function
\be
\Psi(s) = \sum_{t=0}^{\infty} s^{-t} \Psi_t
\label{e:07c}
\ee
Thus, we arrive at the spectral problem for $\Psi(s)=(\psi(1,s),..., \psi(x,s),..., \psi(N,s))^{\top}$
\be
\begin{cases}
s\left(\psi(x,s)-\delta_{x,x_0}\right) = \eps_{x-1}\, \psi(x-1,s) + \eps_{x}\, \psi(x+1,s) & \mbox{for $x=1,...,N$} \medskip \\ \psi_t(x=0,s)=\psi_t(x=N+1,s) = 0 \end{cases}
\label{e:08c}
\ee

Supposing, for simplicity, that $x_0=1$, we can write the solution of \eq{e:08c} as
\be
\psi(x=1,s) = \frac{\det \tilde{B}_N(s,x_0=1)}{\det \tilde{A}_N(s)}
\label{e:09c}
\ee
where $\tilde{A}_N(s)$ and $\tilde{B}_N(s, x_0=1)$ are, correspondingly,
\be
\tilde{A}_N(s) = \left(\begin{array}{ccccc}
s & -\eps_1 & 0 & \cdots & 0 \smallskip \\  -\eps_1 & s & -\eps_2 & & \smallskip \\  0 & -\eps_2 & s & & \smallskip \\ \vdots &  &  &  & \smallskip \\ & & & s & -\eps_{N-1} \smallskip \\ 0 & & & -\eps_{N-1} & s \end{array} \right); \quad \tilde{B}_N(s) = \left(\begin{array}{ccccc}
s & -\eps_1 & 0 & \cdots & 0 \smallskip \\  0 & s & -\eps_2 & & \smallskip \\  0 & -\eps_2 & s & & \smallskip \\ \vdots &  &  &  & \smallskip \\ & & & s & -\eps_{N-1} \smallskip \\ 0 & & & -\eps_{N-1} & s \end{array} \right)
\label{e:10c}
\ee
Recall that $\psi(x=1,s)\equiv \psi(s)$ is the partition function of trajectories which begin at $x_0=1$ and return to the starting point at the very last step. The total number of steps, $N$, in the paths is controlled by the "fugacity" $s$ in the grand canonical ensemble. The expression for $\psi(s)$ is:
\be
\psi(s) =  \frac{s\det \tilde{A}_{N-1}(s)}{\det \tilde{A}_{N}(s)}
\label{e:11c}
\ee
where $\tilde{A}_{N}(s)$ is given by \eq{e:10c}. The distribution function $\psi(s)$ is still a random variable since matrices $\tilde{A}_{N-1}(s)$ and $\tilde{A}_N(s)$ depend on the disorder. However, due to a specific block-diagonal form of the transfer matrix, \eq{e:11c} can be essentially simplified. Namely, $\psi(s)$ reads
\be
\psi(s) = \frac{s\det \tilde{A}^{reg}_{h-1}(s)}{\det \tilde{A}^{reg}_{h}(s)}
\label{e:12c}
\ee
where $\tilde{A}^{reg}_{h}(s)$ is a "regular" three-diagonal matrix of size $h$ without any disorder. The $h\times h$ matrix $\tilde{A}^{reg}_{h}(s)$ is such that its all off-diagonal elements $\{\eps_{N-1-h},...,\eps_{N-1}\}$ are definitely equal to 1. The eigenvalues $\lambda_{k,h}$ of $\tilde{A}_D^{reg}(s)$ are
\be
\lambda_{k,h} = s+2\cos\frac{\pi k}{h+1}; \qquad (k=1,...,h)
\label{e:13c}
\ee
Substituting \eq{e:13c} into \eq{e:12c} and performing averaging over the distribution of the disorder, $Q(h)=p^h$ we arrive at the following expression for the averaged return partition function:
\be
\la \psi(s) \ra = \sum_{h=1}^{N} p^h \sum_{k=1}^{h}\frac{\sin^2\frac{\pi k}{h+1}}{s+2\cos\frac{\pi k}{h+1}}
\label{e:14c}
\ee
One can straightforwardly extract the leading asymptotics of $\la \psi(s) \ra$ noting that the dominant contributions to \eq{e:14c} come from terms in the sum with a nullifying denominator $\lambda_{k,h}$. Equation $\lambda_{\bar{h},\bar{h}}=0$ at $\bar{h}\gg 1$, reads $\disp s-2+\frac{\pi}{\bar{h}^2}=0$, which gives $\bar{h}=\pi/\sqrt{2-s}$. Thus near the spectral boundary $s=2$ the leading exponential behavior of $\la \psi(s) \ra$ is
\be
\la \psi(s) \ra \approx p^{\pi/\sqrt{2-s}} = e^{\pi \ln p/ \sqrt{2-s}}
\label{e:15c}
\ee

\subsubsection{Fine structure of spectral statistics}

\begin{figure}[ht]
\includegraphics[width=16cm]{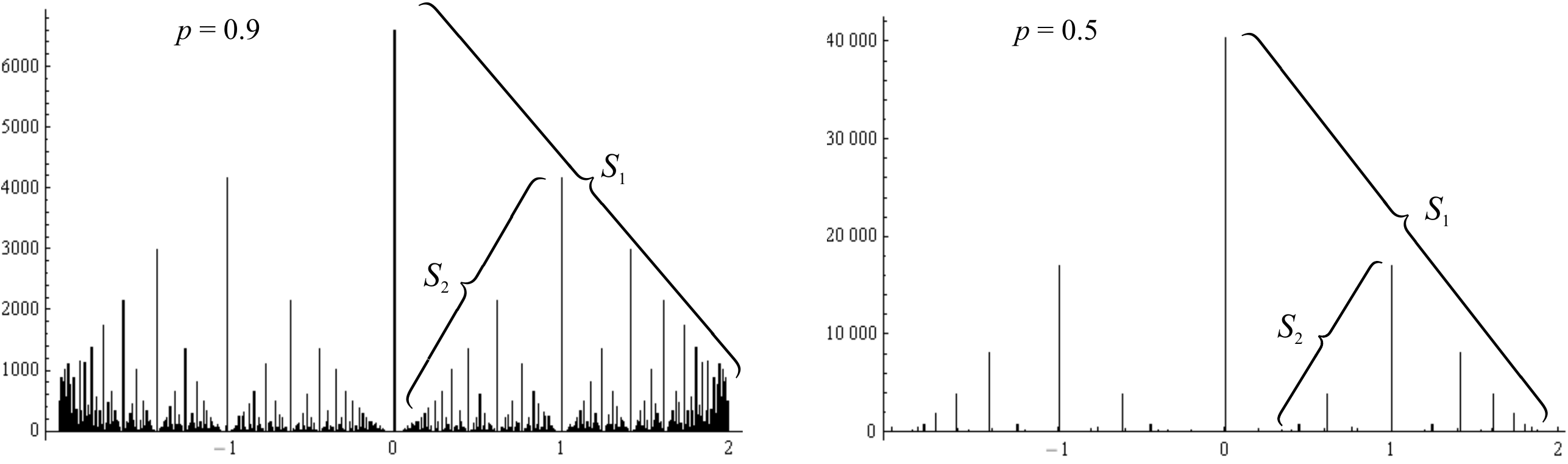}
\caption{The spectral density $\rho(\lambda)$ for the ensemble of three-diagonal random operators of size $N=200$ at $p=0.9$ (a) and $p=0.5$.}
\label{fig:05}
\end{figure}

The sample plots $\rho(\lambda)$ for two different values of $p$, namely for $q=0.9$ and
$p=0.5$ computed numerically for $N=200$ and 500 different realizations, are shown in
the \fig{fig:05}.

The set of eigenvalues of a symmetric gapless $h\times h$ three-diagonal block $A_h$ with $x_k=1$ for all $k=1,...,h$, is given by \eq{e:13c}. The spectral density $\rho(\lambda)$ of the ensemble of $N\times N$ random matrices $A_N$ with the bimodal distribution of matrix elements can be written as a resolvent:
\be
\rho(\lambda) = \lim_{N\to\infty}\frac{1}{N}\la \sum_{n=1}^N\sum_{k=1}^{n} \delta(\lambda-\lambda_{k,h}) \ra =  \lim_{\stackrel{N\to\infty}{\eps\to 0}} \frac{\eps}{\pi N} \sum_{h=1}^N Q_h \sum_{k=1}^h \im\, \frac{1}{\lambda-\lambda_{k,h} - i\eps}
\label{eq:37}
\ee
where $\la ...\ra$ means averaging over the distribution $Q_h=p^h$, and we have used the identity
\be
\delta(\lambda) =\frac{1}{\pi} \lim_{\eps\to+0} \im \frac{\eps}{\lambda- i \eps}
\label{eq:38}
\ee
Substituting \eq{e:13c} into \eq{eq:37}, we find an expression for the density of eigenvalues, $\rho(\lambda)$ of ensemble of random $N\times N$ matrices $A_N$ with the Bernoullian distribution of matrix elements:
\be
\rho(\lambda) = \lim_{\stackrel{N\to\infty}{\eps\to 0}} \frac{1}{\pi N} \sum_{h=1}^{N} p^h
\sum_{k=1}^h\frac{\eps}{\left(\lambda-2\cos\frac{\pi k}{h+1}\right)^2+\eps^2}
\label{eq:39}
\ee

The sum in \eq{eq:39} looks rather complicated, however it is still possible to evaluate in closed form the spectral density, $\rho(\lambda)$, in the limit $N\to\infty$. The resulting expression reads
\be
\rho(\lambda) = \frac{p^{1/g(u)}} {1-p^{1+1/g(u)}}; \qquad u= \frac{1}{\pi}\arccos\frac{\lambda}{2}
\label{eq:40}
\ee
where $g(x)$ is the Riemann function defined in \eq{eq:23}.

\begin{figure}[ht]
\includegraphics[width=16cm]{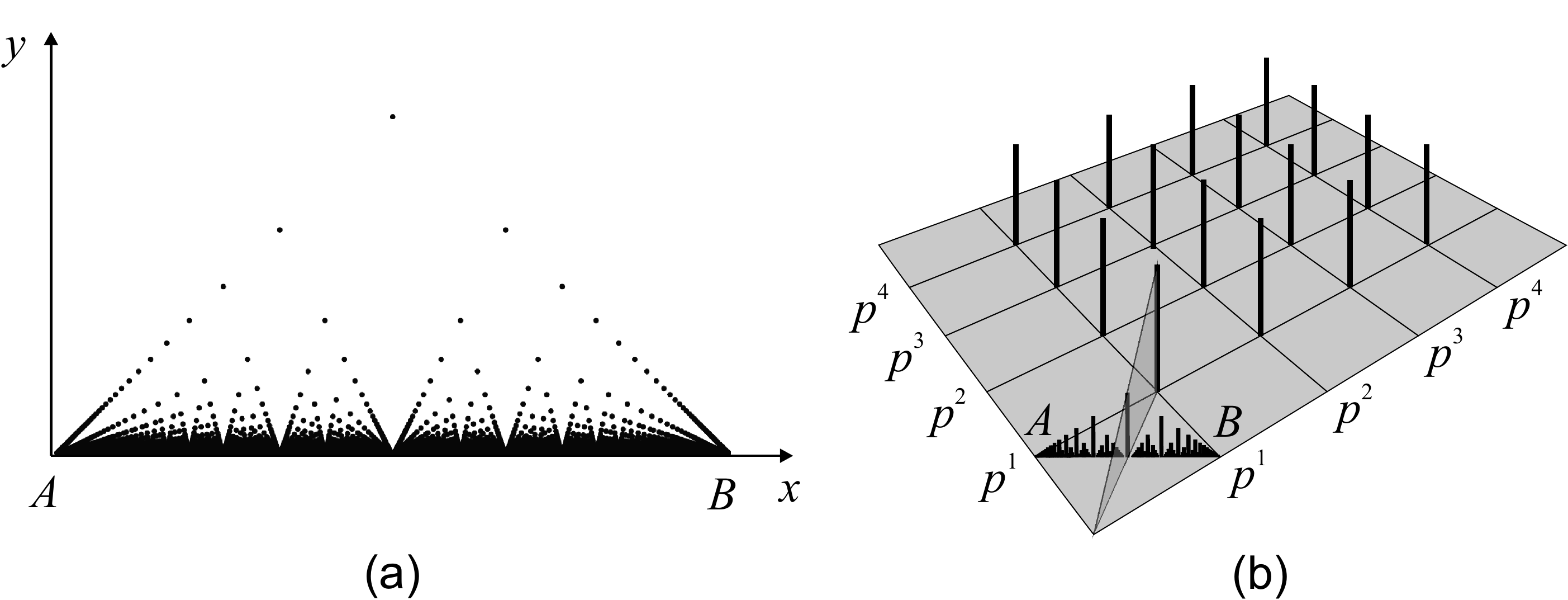}
\caption{(a) Riemann function, (b) Riemann function constructed by the Euclid Orchard.}
\label{fig:03}
\end{figure}

The Riemann function \cite{riemann}, $g(x)$, known also as the Thomae function, has many other names: the popcorn function, the raindrop function, the countable cloud function, the modified Dirichlet function, the ruler function, etc. It is one of the simplest number-theoretic functions possessing nontrivial fractal structure (another famous example is the everywhere continuous but nowhere differentiable Weierstrass function). The Riemann function is defined in the open interval $x \in (0,1)$ according to the following rule:
\be
g(x) = \begin{cases} \frac{1}{n} & \mbox{if $x=\frac{m}{n}$, and $(m,n)$ coprime} \medskip \\
0 & \mbox{if $x$ is irrational} \end{cases}
\label{eq:23}
\ee
The Riemann function $g$ is discontinuous at every rational point because irrationals come
infinitely close to any rational number, while $g$ vanishes at all irrationals. At the same time,
$g$ is continuous at irrationals -- see \fig{fig:03}.

\begin{figure}[ht]
\includegraphics[width=16cm]{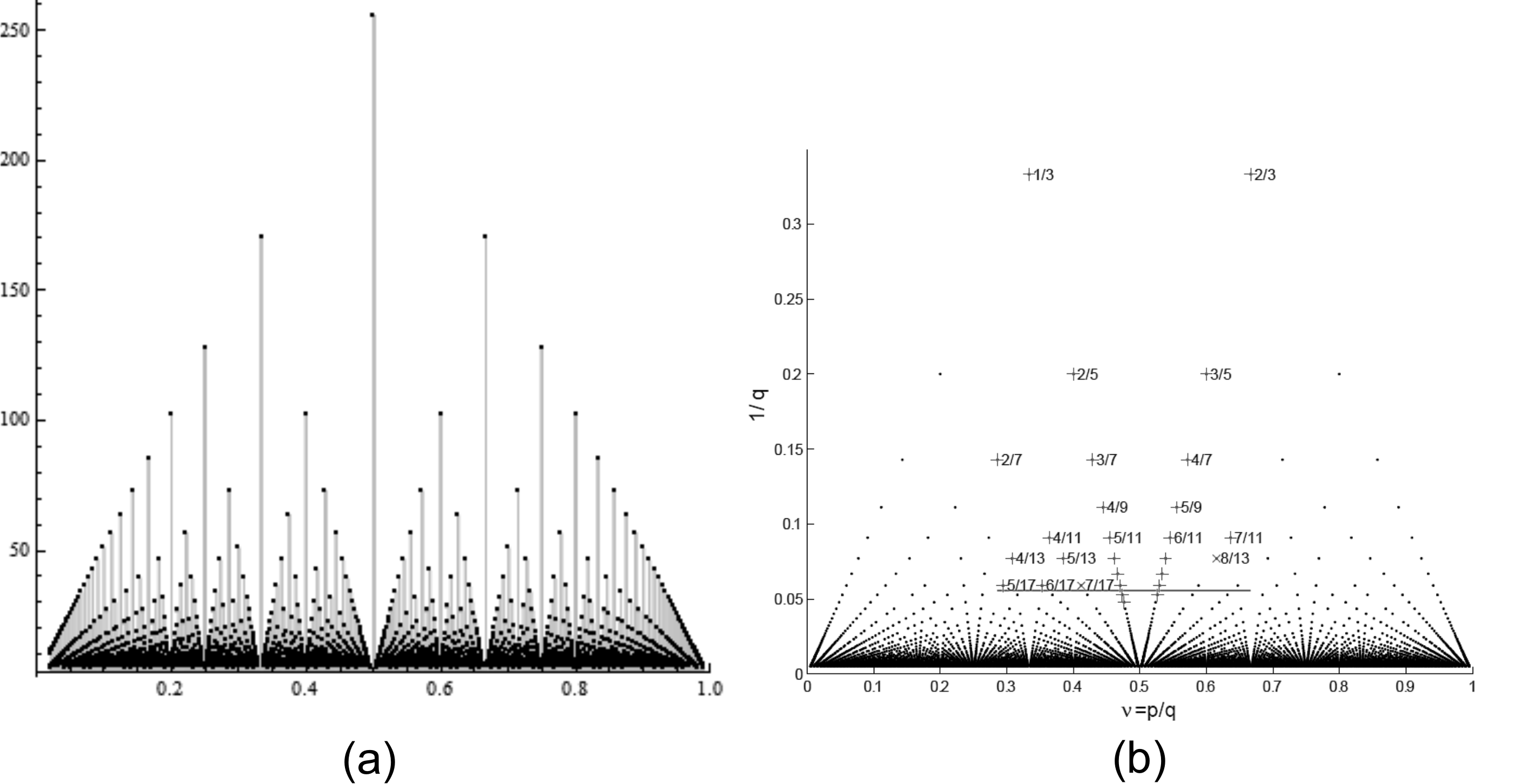}
\caption{(a) Regularization of the normalized everywhere discontinuous Riemann function $f_1(x) = \left(\frac{\pi}{12\eps}\right)^{1/2} g(x)$ (points) by the everywhere continuous function $f_2(x) = \sqrt{-\ln|\eta(x+iy)|}$ (lines) for $y = 10^{-6}$ in the interval $0<x<1$; (b) The stability diagram of Tao-Thouless Fractional Quantum Hall states reproduced from \cite{bergholz}.}
\label{fig:04}
\end{figure}

It has been shown in \cite{polov} that the Riemann function can be regularized as follows
\be
g(x) \to \sqrt{-\frac{12y}{\pi} \ln|\eta(x+i y)|}\Bigg|_{y \to 0}
\label{eq:28}
\ee
where
\be
\eta(z)=e^{\pi i z/12}\prod_{n=0}^{\infty}(1-e^{2\pi i n z})
\ee
is the Dedekind $\eta$-function. The argument $z=x+iy$ is called the modular parameter and $\eta(z)$ is defined for $y>0$ only. The plot of the Riemann function $g(x)$ defined in \eq{eq:23}, as well as its regularization \eq{eq:28} at small $y$ are shown in \fig{fig:04}a, while in \fig{fig:04}b we have reproduced the stability diagram of Tao-Thouless Fractional Quantum Hall states obtained in \cite{bergholz, bergholz2}. The $x$-variable in \eq{eq:28} match the QH filling fractions $\nu=\frac{p}{q}$ drawn along the horizontal axis in \fig{fig:04}b, while the $y$-variable in \eq{eq:28} is the "resolution" of \fig{fig:04}a: as smaller $y$, as larger is $q$ and as more peaks are seen.

\section{Stretched 2D Brownian excursions above impenetrable circular voids}
\label{app2}

\subsection{Model A: Stretched Brownian excursions above semi-circle}

We are interested in the partition function of the 2D Brownian excursion avoiding the disc of the radius $R$. Due to the symmetry of the problem, the setting of the model as shown in \fig{fig:AB}a: the path begins at the point $A(\rho_A, \phi_A)$, runs $N$ steps above the disc till the point $O(\rho,\phi)$, and then, after $N$ more steps ends at the point $B(\rho_B, \phi_B)$. Define by $P_N(\rho,\phi|\rho_A,\phi_A)$ and $P_N(\rho,\phi|\rho_B,\phi_B)$ the partition functions of the Brownian excursion parts $A\to O$ and $O\to B$  correspondingly. Then the partition function of our interest is
\be
P_N(\rho,\phi) = \frac{P_N(\rho,\phi|\rho_A,\phi_A) P_N(\rho,\phi|\rho_B,\phi_B)}{P_N(\rho_A,\phi_A|\rho_B,\phi_B)}
\label{e:01}
\ee
Using $P_N(\rho,\phi)$ we can compute how the Brownian excursion deviates from the impenetrable boundary of the disc. Let us fix some value $\phi=\phi_O$ and compute the expectation and the variance of the distribution function:
\be
\disp \la \rho\ra = \frac{\int_R^{\infty} \rho P_N(\rho,\phi) \rho d\rho}{\int_R^{\infty} P_N(\rho,\phi) \rho d\rho}; \qquad \la \rho^2\ra = \frac{\int_R^{\infty} \rho^2 P_N(\rho,\phi) \rho d\rho}{\int_R^{\infty} P_N(\rho,\phi) \rho d\rho}
\label{e:02}
\ee

The new crucial ingredient of our setting is that we consider the Brownian excursion in a specific "stretched" regime, i.e. we impose the condition $N=cR/a$, where $c$ is some constant. Certainly, $c$ is such that the total length of the Brownian excursion, $2Na$, should be larger than the corresponding arc length, $|\phi_A-\phi_B|R$. We compute the scaling exponent $\gamma$ in the dependence
\be
\mathrm{Var}[r(N)]\equiv\la r^2 \ra-\la r \ra ^2 \sim N^{\gamma}
\ee
at $N\gg 1$ and show that there is a phase transition in $c$: at sufficient stretching (i.e. at $c<c_{tr}$) the typical deviation of the 2D Brownian path from the disc boundary is controlled by the Kardar-Parisi-Zhang exponent, $\gamma = \frac{1}{3}$; while at weak stretching (i.e. at $c>c_{tr}$) the boundary effects vanish and paths statistics becomes Gaussian with $\gamma= \frac{1}{2}$. The transition is very smooth and the qualitative arguments based on the "geometric optic approach" \cite{baruch} permit to conjecture that it is of the third order.

The partition function $P_N(\rho,\phi|\rho_A,\phi_A)$ of the Brownian excursion $A\to O$ satisfies the diffusion equation in polar coordinates:
\be
\begin{cases}
\partial_N P_N(\rho,\phi|\rho_A,\phi_A) = D\Delta_{\rho,\phi}P_N(\rho,\phi|\rho_A,\phi_A) \medskip \\ P_N(\rho=R,\phi|\rho_A,\phi_A))= P_N(\infty,\phi|\rho_A,\phi_A)= P_N(\rho,0|\rho_A,\phi_A)=P_N(\rho,2\pi|\rho_A,\phi_A)=0 \medskip \\
P_N(\rho,\phi|\rho_A,\phi_A)=\frac{1}{\rho}\delta(\rho-\rho_A)\delta(\phi-\phi_A)
\end{cases}
\label{e:03}
\ee
Here $D=\frac{a^2}{4}$ is the two-dimensional diffusion coefficient. The computation of the partition function of the Brownian excursion making the full turn around the impenetrable disc, is a hard technical problem due to the slow convergence of numeric results. To obtain the results with the reliable precision, we have considered Brownian excursions starting at $A(\rho=R+\eps, \phi_A=-\pi)$ and ending at $B(\rho=R+\eps, \phi_B=0)$, and the fluctuations are measured at the middle point point $O(r=\rho-R,\phi_O=\frac{\pi}{2})$. Such a setting differs in details from the full-turn Brownian excursion, however in the regime of strong stretching provides the similar statistical behavior for the fluctuations. The solution of \eq{e:03} is:
\be
P_N(\rho,\phi|\rho_A,\phi_A)=\frac{2}{\pi}\sum\limits_{k=1}^{\infty} \sin k \phi_A \sin k\phi\int\limits_{0}^{\infty}\exp(-\lambda^2 DN)Z_k(\lambda\rho,\lambda R)Z_k(\lambda\rho_A,\lambda R)\lambda d\lambda
\label{e:04}
\ee
where the function $Z_k$ is determined as follows:
\be
Z_k(\lambda\rho,\lambda R)=\frac{-J_k(\lambda\rho)N_k(\lambda R)+J_k(\lambda R)N_k(\lambda\rho)}{J^2_k(\lambda R)+N^2_k(\lambda R)},
\label{e:05}
\ee
with the orthogonality conditions
\be
\int\limits_{0}^{\infty}Z_k(\lambda\rho,\lambda R)Z_k(\lambda\rho_0,\lambda R)\lambda d\lambda=\frac{\delta(\rho-\rho_0)}{\rho}
\label{e:06}
\ee
Changing in \eq{e:04} the variables $\lambda=\frac{\mu}{R}$ and $\rho=R+r$, we get:
\begin{multline}
P_N(\rho,\phi|\rho_A,\phi_A)=\frac{2}{\pi R^2}\sum\limits_{k=1}^{\infty} \sin k \phi_A\sin k\phi \times  \\ \int\limits_{0}^{\infty}\exp\left(-\frac{\mu^2 DN}{R^2}\right)Z_k\left(\mu+\mu \frac{r}{R},\mu\right)Z_k\left(\mu+\mu \frac{r_A}{R},\mu\right)\mu d\mu
\label{e:07}
\end{multline}

\begin{figure}[ht]
\includegraphics[width=16cm]{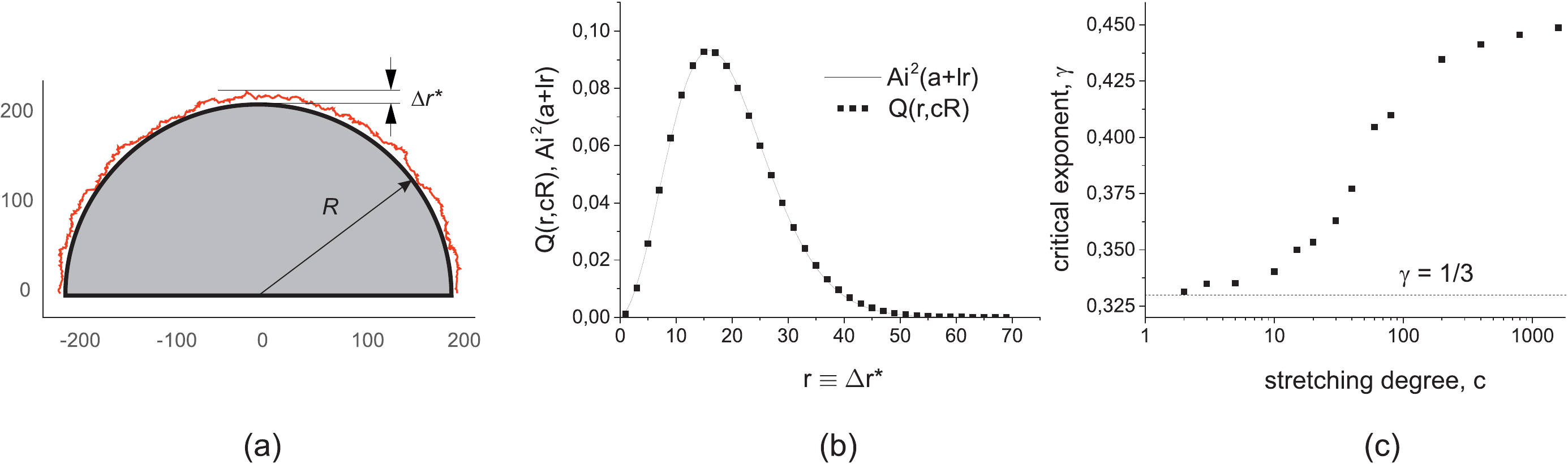}
\caption{(a) Comparison of the distribution $Q(r)$ with $\mathrm{Ai}^2(a_1+a  r)$ for the semicircle radius $R=100$, where $a_1\approx -2.3381$ is the first zero of $\mathrm{Ai}$ and $a \approx 0.08$; (b) Dependence of the critical exponent $\gamma$ on stretching, $c$: at strong stretching ($c\sim \frac{\pi}{2}\div 10$) the exponent $\gamma$ is close to the KPZ exponent $\frac{1}{3}$, at weak stretching $c\gg 1$ the exponent $\gamma$ slowly tends to the Gaussian exponent $\frac{1}{2}$.}
\label{f:01}
\end{figure}

The total probability to find path $A\to O\to B$ at the point $O(r,\phi_O=\frac{\pi}{2})$ above the semicircle can be estimated as
\be
Q(r,N) \propto P_N(\rho,\phi|\rho_A,\phi_A)\times P_N(\rho,\phi|\rho_B,\phi_B) = P_N(\rho,\phi|\rho_A,\phi_A)^2
\label{e:08}
\ee
Recall that we are interested in stretched trajectories only, meaning that we should impose the condition $N=cR$ and consider the typical width of the distribution $Q(r,N)$ defined as follows:
\be
\mathrm{Var}[r(N)]=\int_{0}^{\infty} r^2\, Q\left(r, N=c R\right) dr -\left(\int_{0}^{\infty} r\, Q\left(r,  N=c R\right) dr\right)^2
\label{e:09}
\ee
In \fig{f:01} we have drawn (for $D=1$ and $c=5$) the distribution function $Q\left(r, N=c R/a\right)$ of $r$ at fixed $c=5$ and $R=100$ in comparison with the function
\be
W(r)=\frac{1}{{\cal N}}\, \mathrm{Ai}^2(a_1+a  r), \qquad {\cal N} = \int_0^{\infty} \mathrm{Ai}^2(a_1+a  r)dr
\label{e:09a}
\ee
where $\mathrm{Ai}(z)=\frac{1}{\pi} \int_{0}^{\infty} \cos(\xi^3/3+\xi z)\, d\xi$ is the Airy function, $a_1\approx -2.3381$ is the first zero of $\mathrm{Ai(z)}$, and $a $ is some numeric constant. The function $\mathrm{Ai}^2(a_1+a  r)$ perfectly matches the probability distribution $Q\left(r, N=c R\right)$ -- see \fig{f:01}a. Changing the stretching degree, $c=\frac{N}{R}$, in the interval $c\in[c_{min},c_{max}]$, where $c_{min} = \frac{\pi}{2}$ and $c_{max} = 1500$, we change the critical exponent $\gamma$ in the dependence $\mathrm{Var}[r(N)] \sim N^{\gamma}$, as it is shown in \fig{f:01}b.

At small $c$ (for essentially stretched trajectories) the critical exponent $\gamma$ matches the KPZ scaling with $\gamma=\frac{1}{3}$, while at sufficiently large $c$ ($c>1000$) the exponent $\gamma$ slowly tends to the Gaussian exponent $\frac{1}{2}$. The transition is very smooth and the physical arguments \cite{baruch} permit to conjecture that it is of third order, the visible deviations of $\gamma$ from the KPZ exponent $\frac{1}{3}$ begin at $c\gtrsim 10$.

\subsection{Model B: Fluctuations of inflated loops above impenetrable disc}

Consider the ensemble of 2D Brownian loops of length $L=a  N$, enclosing the area $A$. The corresponding partition function reads
\begin{multline}
Z_N(A) = \int D\{{\bf r}\}\exp\left(\frac{1}{a ^2}\int_0^{N} d\tau \Big(\dot{x}^2(\tau)+ \dot{y}^2(\tau)\Big) \right) \\ \times \delta\left(Aa ^{-2}-\frac{1}{2a ^2}\int_0^{N} d\tau\,\Big(x(\tau) \dot{y}(\tau)-y(\tau) \dot{x}(\tau)\Big)\right)
\label{1}
\end{multline}
The  scaling analysis shows that the corresponding exponent in the path integral and the argument of the $\delta$-function are dimensionless.

Let the area $A$ be strongly inflated such that it looks like a circle of radius $R$ with the fluctuating boundary -- see \fig{fig:AB}b (left panel), i.e, $A\approx \pi R^2 + o(R^2)$. Since the "backtracks" of paths in the inflated regime are exponentially suppressed and the speed of a propagating particle is nearly constant, we can write in polar coordinates $(x=r \cos \phi;\; y=r \sin \phi)$:
\be
A =\frac{1}{2}\int (x dy-ydx) = \frac{1}{2}\int_0^{2\pi} d\phi\; {\bf r}^2(\phi)  =\frac{\pi}{N}\int_0^N d\tau\; {\bf r}^2(\tau)
\label{2}
\ee
If $r(\tau)\equiv R$ for any $\tau$, we get the area of the circle, $A_0=\pi R^2$.

Identically, we can control the inflation degree of an ideal Brownian loop by fixing the typical gyration radius, $R_g^2$ defined as follows:
\begin{multline}
R_g^2=\frac{1}{2N^2}\int\limits_0^N \int\limits_0^N d\tau d\tau'({\bf r}(\tau)-{\bf r}(\tau'))^2 = \frac{1}{N}\int\limits_0^N d\tau \mathbf{r}^2(\tau)-\left(\frac{1}{N}\int\limits_0^N d\tau\, \mathbf{r}(\tau)\right)^2 = \\ \frac{1}{N}\int\limits_0^N d\tau\, \mathbf{r}^2(\tau)-{\bf R}_c^2 \equiv \frac{1}{N}\int\limits_0^N dt \mathbf{r}^2(\tau)
\label{3}
\end{multline}
In \eq{3} it is implied that the position of the center of mass is fixed at 0, i.e. ${\bf R}_c=0$. Under that condition, the path integral \eq{1} reads
\be
Z_N(s) = \frac{1}{2\pi}\int_{\infty}^{\infty} e^{i s A a ^{-2}} Z_N(s)
\label{4}
\ee
where
\be
Z_N(s) =\int D\{{\bf r}\}\exp\left(-\int_0^N \left(\frac{1}{a^2}\, \dot{{\bf r}}^2(t) + \frac{\pi s}{Na^2}\, {\bf r}^2(t)\right)dt\right) = \int D\{{\bf r}\} e^{-S}
\label{5}
\ee
The Lagrangian $L$ of the action $S=\int_0^N L(\tau)\, d\tau$ in \eq{5} is defined as
\be
L(\tau)=\frac{1}{a^2}\,\dot{{\bf r}}^2(\tau) + \frac{\pi s}{Na^2}\,{\bf r}^2(\tau)
\label{6}
\ee
and the corresponding nonstationary two-dimensional Schr\"odinger-like equation for the probability distribution in the radial parabolic well $V({\bf r}) = -\frac{\pi s}{Na^2}\,{\bf r}^2$ is
\be
\frac{\partial W({\bf r},\tau)}{\partial \tau} = \frac{a^2}{4} \nabla^2 W({\bf r},\tau) -  \frac{\pi s}{Na^2}\, {\bf r}^2\, W({\bf r},\tau)
\label{7}
\ee
Making in \eq{7} the substitution $W({\bf r},\tau)= U({\bf r},\tau) r^{-1/2}$, we get the equation in the radial framing
\be
\frac{\partial U(r,\tau)}{\partial \tau} = \frac{a^2}{4} \frac{\partial^2 U(r,\tau)}{\partial r^2} + \left(\frac{a^2}{16 r^2} - \frac{\pi s}{Na^2}\, r^2 \right) U(r,\tau)
\label{7a}
\ee
As we shall see below by the self-consistent arguments, in the regime of strong inflation of the loop, we have $\left|\frac{a^2}{16 r^2}\right| \ll \left|\frac{\pi s r^2}{Na^2}\right|$ and hence the equation \eq{7a} ca be written as follows
\be
\frac{\partial U(r,\tau)}{\partial \tau} = \frac{a^2}{4} \frac{\partial^2 U(r,\tau)}{\partial r^2} - \frac{\pi s}{Na^2}\, r^2 U(r,\tau)
\label{7b}
\ee
Separating the variables in \eq{7b}, we get
\be
\begin{cases}
\disp -\lambda_n U_n(r,\tau) = \frac{\partial U_n(r,\tau)}{\partial \tau} \medskip \\
\disp -\lambda_n U_n(r,\tau) = \frac{a^2}{4} \frac{\partial^2 U_n(r,\tau)}{\partial r^2} - \frac{\pi s}{Na^2}\, r^2\, U_n(r,\tau)
\end{cases}
\label{10}
\ee
The solution of the stationary polymer problem \eq{10} is
\be
\begin{cases}
\disp U_n(r) = \left(\frac{4 s}{\pi Na^4}\right)^{1/4}\frac{1}{\sqrt{2^n n!}} H_n\left(r\left(\frac{4 \pi s}{Na^4}\right)^{1/4} \right)\exp\left(-r^2\left(\frac{\pi s}{Na^4}\right)^{1/2}\right) \medskip \\ \disp \lambda_n=\left(n+\frac{1}{2}\right)\sqrt{\frac{\pi s}{N}}
\end{cases}
\label{12}
\ee
where $H(...)$ is the Hermite polynomial.

We are interested in the following distribution function of the Brownian excursion:
\be
Q(r,N) \propto \frac{1}{r} \sum_{n=0}^{\infty}e^{-\lambda_n N}\, U_n^2(r)
\label{13}
\ee
where the eigenfunction $U_n(r)$ and corresponding eigenvalue $\lambda_n$ are defined in \eq{12}. The function $U_n(r)$ defines the probability of the \emph{open end} of a polymer to be at the point $x$, thus the function $U_n^2(r)$ is the Brownian bridge probability, i.e. of the intermediate point of the polymer loop to be at the point $r$.

Let us estimate the sum in \eq{13} via the saddle point method. To proceed, recall that $s$ is the Lagrange multiplier of the inflated area $A$. Thus, to fix "softly" the trajectories with given $A$, we set in \eq{5}
\be
s=a^2 A^{-1}
\label{14}
\ee
($s$ is the dimensionless quantity). For inflated trajectories, which are close to the perfect circle of the radius $R=\frac{Na}{2\pi}$, the area scales as $A=\pi R^2= \frac{N^2a^2}{4}$ and, hence, $s\approx 4N^{-2}$. From \eq{13} we see that the dominant contribution to $Q(x,N)$ comes $n$, such that $\lambda_n\, N \approx 1$. Plugging the expression $\lambda_n = N^{-1}$ into \eq{12}, we arrive at the equations which determines the values of $n$ which give the main contribution to $Q(x,N)$:
\be
\frac{1}{N} = \left(n+\frac{1}{2}\right)\sqrt{\frac{\pi s}{N}}
\label{15}
\ee
Solving \eq{15} at $n\gg 1$, and using \eq{14}, we get
\be
n\equiv n^* \approx \frac{1}{\sqrt{\pi sN}}=\sqrt{\frac{A}{\pi Na^2}}
\label{16}
\ee
Expressing all parameters in \eq{12} in terms of $A$ and $N$, we can rewrite \eq{12} as follows
\be
U_n(x) = \left(\frac{4}{\pi A Na^2}\right)^{1/4}\frac{1}{\sqrt{2^n n!}} H_n\left(r\left(\frac{4\pi}{A Na^2}\right)^{1/4} \right)\exp\left(-r^2\left(\frac{\pi}{A Na^2}\right)^{1/2}\right)
\label{17}
\ee
It is known that the Hermite polynomials $H_n(z)$ at $z\approx \sqrt{2 n}$ and $n\gg 1$ have the following uniform asymptotic expansion
\be
H_n(z)\approx \sqrt{2\pi}\, \exp\left(\frac{n\ln(2n)}{2}-\frac{3n}{2}+z\sqrt{2 n}\right)
n^{1/6}\mathrm{Ai}\left(\sqrt{2}\,\frac{z-\sqrt{2 n}}{n^{-1/6}}\right)
\label{18}
\ee
where $z=r\left(\frac{4\pi}{ANa^2}\right)^{1/4}$. The uniform asymptotics \eq{18} is valid only when $z^*\approx \sqrt{2 n^*}$. This condition fixes the equation for $r=r^*$, at which the Airy tail of the Hermite polynomials appears:
\be
z^*\approx \sqrt{2 n^*} \; \to \; r^*\left(\frac{4\pi}{A Na^2}\right)^{1/4} \approx  \left(\frac{4 A}{\pi Na ^2}\right)^{1/4}
\label{19}
\ee
The argument of the Airy function in \eq{18}, $\disp \frac{z-\sqrt{2 n}}{n^{-1/6}}$, can be rewritten as follows:
\be
\frac{z^*-\sqrt{2 n}}{n^{-1/6}} \equiv \frac{\disp r^*\left(\frac{4\pi}{A Na^2}\right)^{1/4} - \left(\frac{4 A}{\pi Na^2}\right)^{1/4}} {\disp \left(\frac{A}{\pi Na^2}\right)^{-1/12}} = \xi
\label{20}
\ee
where $\xi$ is the numerical value of order of 1. Finding $r^*$ from the solution of \eq{20}, we get
\be
\Delta r^* \equiv r^* -\la r^*\ra = \frac{\xi}{2^{1/2}}\left(\frac{Aa^4}{\pi}\right)^{1/6}N^{1/3}
\label{21}
\ee
where for the mean value $\la r^*\ra$ we have: $\disp \la r^*\ra = \frac{A^{1/2}} {\pi^{1/2}}$.

The cutoff in \eq{13} of modes with small eigenvalues $\lambda_k$ (i.e. large wavelengths) does not permit the inflated ring to possess large-scale fluctuations. Such a cutoff can be ensured by introducing the hard-wall constraint in a form of an impenetrable disc as shown in \fig{fig:AB}b (right panel) which prevents the loop of large-scale fluctuations. In the absence of the hard-wall constraint the fluctuations of the inflated disc are Gaussian which we do see in our numeric simulations.

According to \eq{21} for the typical span of the path's fluctuations, $\Delta r^*$, with respect to the mean value, $\la r^*\ra$, we expect the Kardar-Parisi-Zhang scaling: $\Delta r^* \sim N^{1/3}$, i.e. the path gets localized near the disc boundary within a circular strip of width $\sim N^{1/3}$.

\begin{figure}[ht]
\includegraphics[width=16cm]{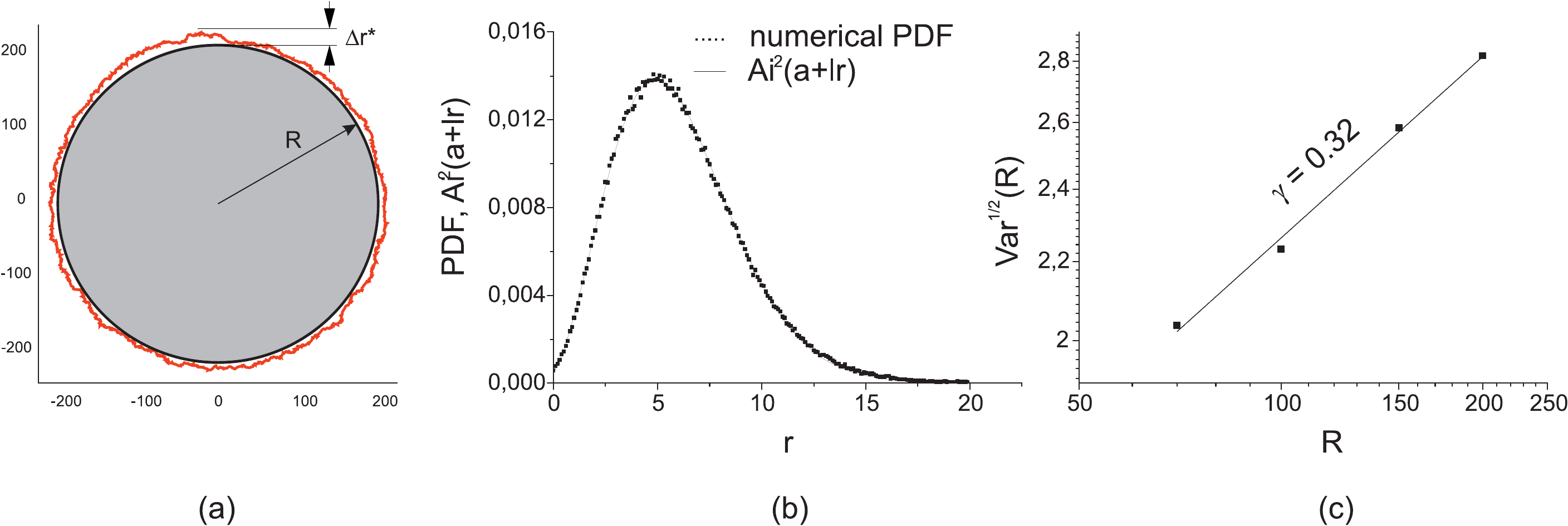}
\caption{(a) Distribution of the fluctuations of the inflated ring above the impenetrable disc; (b) Scaling of fluctuations as a function of the path length in log-log coordinates.}
\label{fig07}
\end{figure}

To check whether the prediction \eq{21} is correct, we have realized the setting of the model B in the numeric simulation. The typical snapshot of the strongly inflated loop above the inserted impenetrable disc is depicted in the right panel of \fig{fig:AB}. In the left panel of \fig{fig07} we have plotted the distribution function of fluctuations $\Delta r^*$ which actually coincide with the square of the Airy function. In the right panel of \fig{fig07} we reproduce the numerically obtained dependence of $\Delta r^*(N)\sim N^{\gamma}$ with $\gamma \approx 0.32$, which is very close to the KPZ critical exponent $\gamma=\frac{1}{3}$.

\begin{figure}[ht]
\includegraphics[width=16cm]{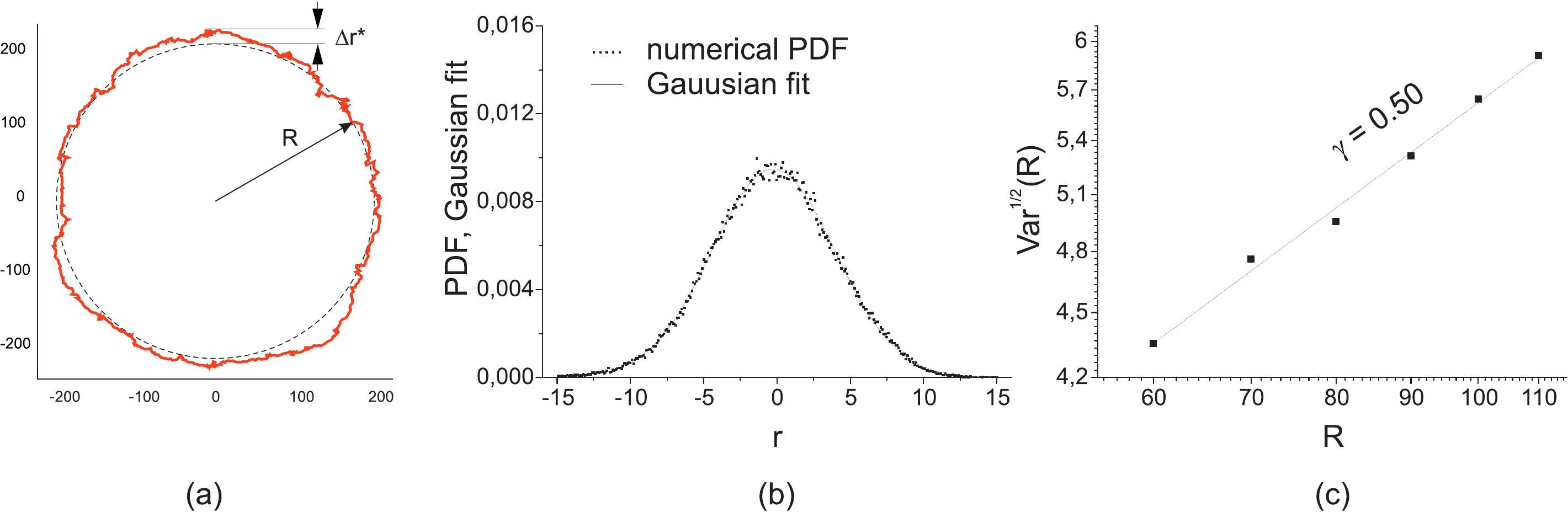}
\caption{(a) Distribution of the fluctuations of the inflated ring without the boundary; (b) Gaussian scaling of fluctuations as a function of the path length in log-log coordinates.}
\label{fig08}
\end{figure}

We argue that the simultaneous fulfilment of two conditions: (i) the path stretching and (ii) the hard-wall curved constraint, restricting the large-scale fluctuations, is crucial for the localization of paths within the strip of width $N^{1/3}$. By the path stretching the trajectories are pushed to an improbable tiny region of the phase space, however the presence of a large deviation regime is not sufficient to affect the path's  statistics, and the presence of the solid curved boundary on which paths are leaning, is crucial. The importance of a solid convex boundary has been studied in \cite{peres,shlosman}. In \fig{fig08} we have shown the distribution function of monomers of inflated loop and variance of their fluctuations. One can clearly seen that in absence of the solid boundary the fluctuations are Gaussian.

Using \eq{21} we can straightforwardly estimate the free energy of the path confined in a circular strip of width $\disp \Delta r^*(N)  = \frac{\xi}{2^{1/2}}\left(\frac{Aa^4}{\pi}\right)^{1/6} N^{1/3}$. Proceeding as in \eq{eq:f(f)}, the corresponding free energy of the ensemble of long ($N\gg 1$) paths can be written as follows:
\be
F(N) = N \lambda_{min}
\label{fr-KPZ}
\ee
where $\lambda_{min}$ is the smallest eigenvalue of a randomly fluctuating Brownian motion within the strip of width $\Delta r^*$, i.e. $\lambda_{min} = \frac{2\pi a^2}{(\Delta r^*)^2}$. Plugging the explicit expression of $\Delta r^*(N)$ into $\lambda_{min}$, we get for the free energy $F(N)$:
\be
F(N) = \frac{2\pi Na^2}{(\Delta r^*)^2} \approx \frac{4\pi}{\xi^2} \left(\frac{a}{R}\right)^{2/3} N^{1/3}
\label{free-KPZ-2}
\ee
In the last expression we have supposed that the strongly inflated ring encloses the area $A\approx \pi R^2$. The corresponding Gibbs measure reads
\be
W(N) \propto \exp(-F(N)) \approx \exp\left(-\alpha \varkappa^{2/3} N^{1/3}\right)
\label{gibbs2}
\ee
where $\alpha\frac{4\pi}{\xi^2}$ is the numeric model-dependent coefficient and $\varkappa=\frac{a}{R}$ is the curvature of the bounding disc.

It is worth noting that the curvature, $\varkappa$, enters into the survival probability \eq{gibbs2} in the same way as the strength of the Poissonoan disorder, $\beta$, enters into the survival probability \eq{eq:surv} for $N$-step paths in the Poissonian distribution of traps on the line. Found similarity between \eq{gibbs2} and \eq{eq:surv} permits us to claim that the interplay of the curvature and the stretching (or, in terms of the work \cite{stanford}, of the external pressure), is the source of the emergence of the Griffits-like behavior \eq{gibbs2}, which is based upon the KPZ-like fluctuations.

\subsection{Scaling analysis of stretched paths in Poissonian disorder}

We have argued above that the LT for Gaussian disorder in the boundary theory can be captured
holographically via the particular limit of the trajectories of the particle at the hyperbolic plane in effective external magnetic field whose strength corresponds to the strength of the
boundary disorder. Here we explain which type of trajectories are relevant for the LT in the Poissonian disorder at the boundary.

We study a two-dimenional model possessing the scaling as is \eq{eq:surv} in a large-deviation regime. We are inspired by the (1+1)D model of H. Spohn and P. Ferrari \cite{ferrari} where the statistics of 1D directed random walks evading the parabolic void has been discussed. Such a system was proposed as a simplified "mean-field-like" model describing the fluctuations of a top line in a bunch of $n$ 1D directed "vicious walks" (world lines of free fermions in 1D). Proceeding as in \cite{ferrari}, we can define the averaged position of the top line and look at its fluctuations. In such a setting, all vicious walks lying below the top line, play a role of a "mean field", pushing the top line away from the bulk to some new equilibrium position. Replacing the effect of the "bulk" by the semicircle, one arrives at the Spohn-Ferrari model where the 1D directed random walk stays above the semicircle, and its interior is inaccessible for the path. In \cite{ferrari} the authors have shown that the (1+1)D system has fluctuations controlled by the KPZ exponent, $\nu=\frac{1}{3}$. In the work \cite{valov}, the 2D random walk evading the semicircle was considered in a special regime, in which the path length is comparable to the length of the semicircle arc. Such a "stretching" condition forces the random walk to stay near the semicircle surface and influences drastically the typical path span above the curved surface.

\begin{figure}[ht]
\includegraphics[width=16cm]{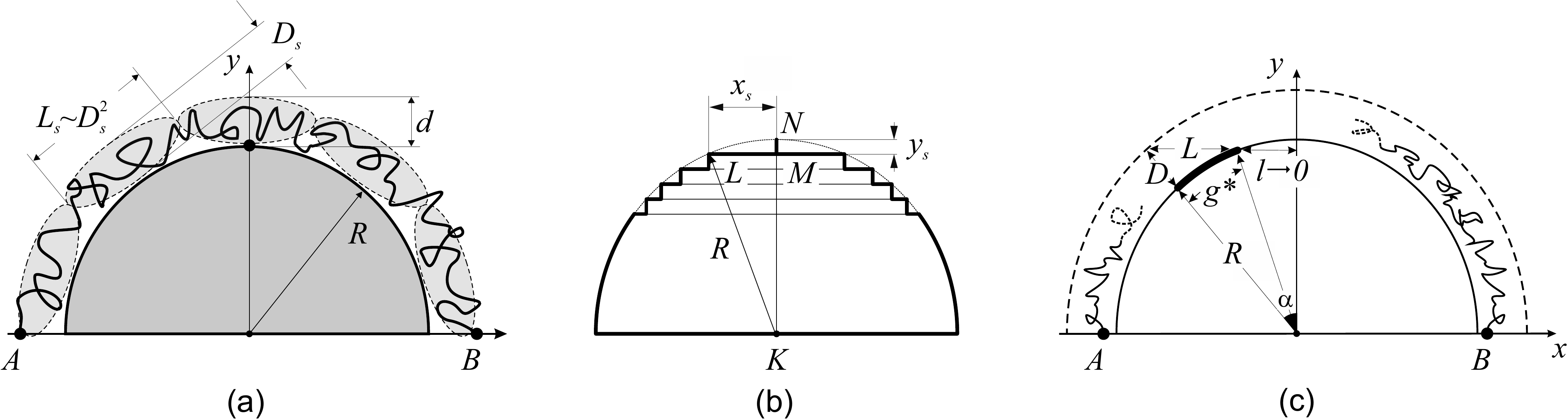}
\caption{(a) Two-dimensional random walk evading the semicircle. Dashed "elliptic blobs" of longitudinal size $L_s \sim D_s^2/a$ designate uncorrelated parts of a path; (b-c) Auxiliary geometric constructions for \eq{e:30} and \eq{e:36}.}
\label{fig:01}
\end{figure}

We consider an ensemble of 2D random $N$-step paths (each step has length $a$) stretched above a semicircle of radius $R$. Stretching is ensured by the restriction on a path length, $L=aN$, such that $c\left(\frac{R}{a}\right) < N \ll \left(\frac{R}{a}\right)^2$ where $c$ is some constant. In what follows we set $a=1$. Schematically, the system under consideration is shown in \fig{fig:01}.

The resulting paths configurations are "atypical" since their realizations would be highly improbable in the ensemble of unconstrained Gaussian trajectories. Statistics in such a tiny subset of a Gaussian ensemble is controlled by a collective behavior of strongly correlated modes, thus, for some geometries one might expect extreme distribution with a scaling different from the Gaussian. Let us briefly reproduce the arguments of the work \cite{valov}. Typically, an elongated path follows the stretching direction as much as possible, and gets curved only if curving cannot be avoided. To make our arguments as transparent as possible, consider \fig{fig:01}b and denote by $y_s$ an average span of the path in vertical direction above the tip of the semicircle, and by $x_s$ -- the typical size of the horizontal segment, along which the semicircle can be considered as nearly flat. Thus, above the segment $LN$ the path statistics is not perturbed by the curvature of the semicircle boundary. In the limit $y_s \ll R$, the horizontal segment $LN$ linearly approximates the corresponding arc of the circle as it is shown in \fig{fig:01}b. We have
\be
x_s = \sqrt{R^2 - (R-y_s)^2}\,\Big|_{y_s\ll R} \approx \sqrt{2Ry_s}
\label{e:29}
\ee
Since the horizontal support, $x_s$, is almost flat, the span of the path, $y_s$, in the vertical direction is the same as for an ordinary random walk, i.e. $y_s\sim \sqrt{x_s a}$. This expression together with \eq{e:29} leads us to a self-consistent equation: $x_s \sim \sqrt{R \sqrt{x_s a}}$, which immediately gives
\be
x_s \sim R^{2/3} a^{1/3}; \quad y_s \sim \sqrt{x_s a} \sim R^{1/3} a^{2/3}
\label{e:30}
\ee
To estimate the free energy, $F(N)$, of ensemble of $N$-step paths stretched above the semicircle, note that we can split the entire stretched $N$-step path running from $A$ to $B$ above the semicircle in the sequence of independent "renewal times", shown in \fig{fig:01}a by dashed lines, with the longitudinal size $L_s \equiv x_s \sim R^{2/3} a^{1/3}$ and the width $D_s \equiv y_s \sim R^{1/3} a^{2/3}$.

Taking into account the additivity of the free energy, $F(N)$, we can estimate $F(N)$ of ensemble of stretched $N=cR/a$ ($R=Na/c$)--step paths as
\be
F(N) \sim \frac{Na}{L_s} \sim \frac{Na^2}{D_s^2} \sim  c^{2/3}\,N^{1/3}
\label{e:31}
\ee
If each step of the random walk experiences the action of an external potential field $u$ (for example, an electric field), then the free energy \eq{e:31} acquires the extra term $F_0\sim Nu$, which gives
\be
F(R) \sim F_0 + c^{2/3} N^{1/3} \sim  u\,N + c^{2/3}\,N^{1/3}
\label{e:31a}
\ee
Thus, stretching in curved geometry results in splitting a chain in a sequence of "renewal times" $t_{opt}\sim  L_s/a \sim N^{2/3}$ and hence, to the emergence of an intrinsic scale of order of order of $t_{opt}$. The Gibbs measure, $P(N)$, which provides the free energy distribution of stretched trajectories of length $N=cR/a$ in the curved channel of width $D_s$ and the drift field $u$, can be straightforwardly estimated using \eq{e:31a}:
\be
P(N) = \exp(-F(N)) \sim \exp\left(-u\, N - c^{2/3}\,N^{1/3}\right)
\label{e:33}
\ee
Comparing \eq{e:33} and \eq{eq:surv} we see that the $P(N)$ and $W(N)$ look almost identical, besides the finite size corrections controlled by the critical exponent $\nu=\frac{1}{3}$ have different meaning: they are induced by the disorder in \eq{eq:surv} and by large deviations in curved geometry in \eq{e:33}. The Legendre transformation in \eq{e:33} from $N$ to the conjugated variable $E$, realized via the inverse Laplace transform as in \eq{e:03a}, gives the Lifshitz tail
\be
P(E) \propto \frac{1}{2\pi i}\int_{\eps-i\infty}^{\eps+i\infty} P(N)\, e^{NE} dR\propto
\exp\left(-\frac{c}{\sqrt{E-u}}\right)
\label{eq:34}
\ee

The method of optimal fluctuations directly applied to stretched paths above the disc permits us to derive $F(N)$ in \eq{e:33}. Following the way of reasoning as in \eq{eq:surv}, let us write the expression for the nonequilibrium free energy $F(D_s,R)$ of the stretched ($N=c R/a$) paths confined in the curved slit of width $D_s$
\be
F(D_s,R)\sim \frac{Ra}{D_s^2}-\frac{R}{a t_{opt}(R)}
\label{e:35}
\ee
where the first term encounters the entropic loss due to the path's compression and the second term is the number of independent renewal times in the curved slit (by $t_{opt}(R)$ we denote the average number of chain monomers in the typical "renewal time"). To find the equilibrium value $\bar{D_s}(R)$ we should minimize $F(D_s,R)$ with respect to $D_s$.

The typical "renewal time" during which the path does not touch the boundaries of a channel, determines the width of a channel. In \fig{fig:01}c we have denoted this part by $\bar{L}$. A simple geometric analysis allows us to express $\bar{L}$ as a function of $R$, $D_s$ and $\alpha$:
\be
\bar{L}=(R+D_s)\sin\alpha - l = (R+D_s)\sin\alpha -\sqrt{R^2-(R+D_s)^2\cos^2\alpha}
\label{e:36}
\ee
Note now that the angle $\alpha$ cannot be less than $\alpha_{min}$ defined by the condition $l(\alpha_{min})=0$. The value of $\alpha_{min}$ in the limit $R\gg D_s$ can be easily estimated: $\alpha_{min} \approx \sqrt{\frac{D_s}{R}}$. Thus, typical $t_{opt}$ is then $t_{opt} \approx R\alpha_{min}/a$, which gives the following expression for $F(D_s,R)$:
\be
F(D_s,R) \approx \frac{Ra}{D_s^2} - \sqrt{\frac{R}{D_s}}
\label{e:37}
\ee
Minimizing \eq{e:37} with respect to $D_s$, we get the equilibrium $D_s=\tilde{D}_s$ for stretched paths evading the semicircle:
\be
\bar{D_s} \sim R^{1/3}a^{2/3}
\label{e:38}
\ee
which coincides with $y_s$ obtained by hand-waving arguments in \eq{e:30} and provides correct scaling for the renewal time $t_{opt} \sim (R/a)^{2/3}$.

Despite of striking similarity, there is a difference between two methods of optimal fluctuations \eq{eq:free} and \eq{e:37}. The optimal fluctuation for 1D trapping in a disorder deals with balancing the confinement free energy $\propto \frac{Na^2}{\bar{L}^2}$ with the entropy $\sim -\ln Q(\bar{L})$ of Poissonian disorder. However, for the localization of stretched ($N=cR/a$) paths, evading the semicircle of radius $R$, the same confinement free energy, $\sim \frac{Na^2}{D_s^2}$, is balanced with the energy $\propto \sqrt{\frac{R}{D_s}}$ of collection of independent blobs, emerging due to the geometrical constraints in curved geometry.

We can speculate about more physically justified interpretation of these "renewal times" and question if the set of renewal times is the counterpart of the linear structure with the heavy insertions in the boundary theory. The values $t_{opt}$ emerge from defects in the quantum gravity. Recently it was recognized that in the problem of the Page curve for black hole, the replica wormholes has to be taken into account \cite{shenker, maldacenaworm}. The replica wormhole can be represented as the point-like heavy defects in the single copy of geometry and they cure the late time behavior of correlators via the long tails in the distributions. Presumably the  renewal times are related to these gravitational defects, however that point needs for clarification.

\end{appendix}

\end{document}